\documentclass[preprint]{aastex631}

\newcommand{\bfx}{{\boldsymbol{x}}}
\newcommand{\bfk}{{\boldsymbol{k}}}

\newcommand{\bfb}{{\boldsymbol{b}}}
\newcommand{\bfd}{{\boldsymbol{d}}}

\newcommand{\hatd}{{\boldsymbol{\widehat{d}}}}
\newcommand{\GHz}{{\text{GHz}}}
\newcommand{\bfnabla}{{\boldsymbol{\nabla}}}
\newcommand{\tmp}{}

\shorttitle{Cosmological Distances to FRBs}
\shortauthors{Boone \& McQuinn}
\graphicspath{{./}{figures/}}

\newcommand{\expectation}[1]{\left\langle #1 \right\rangle}

\usepackage{amsmath}

\begin{document}

\title{Solar System-scale interferometry on fast radio bursts could measure cosmic distances with sub-percent precision}

\author[0000-0002-5828-6211]{Kyle Boone}
\email{kyboone@uw.edu}
\affiliation{Department of Astronomy, University of Washington, 3910 15th Ave NE, Seattle, WA 98195, USA}

\author{Matthew McQuinn}
\email{mcquinn@uw.edu}
\affiliation{Department of Astronomy, University of Washington, 3910 15th Ave NE, Seattle, WA 98195, USA}



\begin{abstract}

The light from an extragalactic source at a distance $d$ will arrive at detectors separated by $100~$AU at times that differ by as much as $120\,(d/100\,{\rm Mpc})^{-1}~$nanoseconds because of the curvature of the wave front. At gigahertz frequencies, the arrival time difference of a point source can be determined to better than a nanosecond with interferometry. If the space-time positions of the detectors are known to a few centimeters, comparable to the accuracy to which very long baseline interferometry baselines and global navigation satellite systems (GNSS) geolocations are constrained, nanosecond timing would allow competitive cosmological constraints. We show that a four-detector constellation at Solar radii of $\gtrsim 10\,$AU could measure geometric distances to individual sources with sub-percent precision. The precision increases quadratically with baseline length. Fast radio bursts (FRBs) are the only known bright extragalactic radio source that are sufficiently point-like for this experiment, and the simplest approach would target the population of repeating FRBs. Galactic scattering limits the timing precision at $\lesssim3\,$GHz, whereas at higher frequencies the precision is set by removing the differential dispersion between the detectors.  Furthermore, for baselines greater than $100~$AU, Shapiro time delays limit the precision, but their effect can be cleaned at the cost of two additional detectors. Outer Solar System accelerations that result in $\sim$1~cm uncertainty in detector positions could be corrected for with weekly GNSS-like trilaterations between members of the constellation. The proposed interferometer would not only provide a geometric constraint on the Hubble constant, but also could advance Solar System, pulsar, and gravitational wave science.


\end{abstract}



\section{Introduction}

Fast radio bursts (FRBs) are millisecond radio transients that are generally of extragalactic origin \citep{lorimer07, thornton13, tendulkar17, bannister19, cordes-review, petroff21}. Hundreds of FRBs have been discovered to date, tens of which have been found to be repeating, with $\sim10^4$ on the sky per day above $1 \;$Jy~ms -- a detectable fluence for many radio telescopes \citep{spitler16, chime-catalogue}.  Not only are FRBs interesting for identifying and studying the extreme radiative processes that create them, likely associated with magnetars \citep{2020Natur.587...59B, 2020Natur.587...54C}, but propagation effects that alter the received electromagnetic waves can be used as a tool to study the missing baryon problem, the circumgalactic media of galaxies, and the cosmic reionization history \citep{mcquinn14, 2019ApJ...872...88R, 2019Sci...366..231P, 2021MNRAS.502.5134B, 2021arXiv210714242H}.  

Several studies have also identified potential ways to use FRBs for precision cosmology. The simplest of which is to use that the frequency dependence of the wave front arrival to infer the column of electrons along the sightline and create a Hubble diagram-like electron column versus redshift relation.  While it is unlikely that the host galaxy electron column will be small enough to allow a precise determination of this relation \citep{2020Natur.581..391M}, a way around this uncertainty is to instead use that correlations between sightlines are only sensitive to the cosmological density fluctuations \citep{2015PhRvL.115l1301M}.  This idea would map the fluctuating electron density to constrain cosmological parameters, somewhat analogous to a weak lensing survey.  



 An alternative route to cosmological constraints uses a key attribute of FRBs -- that their arrival time can be measured very accurately.  The point-like nature of FRBs allows coherent timing with precision of better than the inverse of their frequency, a nanosecond at a gigahertz. FRBs could be timed much more precisely than the slowly varying quasars that are used in current strong lensing time delay analyses, which are starting to put competitive constraints on the Hubble constant \citep{li18}.  Unfortunately, uncertainties in the mass modeling of the lens system \citep{2021MNRAS.501.5021K} as well as in the projected mass \citep{1996ApJ...468...17B} may prevent percent-level cosmology with lensing time delays regardless of the timing precision.
One potential way around these limitations involves lensed, repeating FRBs.  
Measuring the $\sim10^{-3}$~s~yr$^{-1}$ evolution in the time between lensed images
due to the evolving redshift of the lens and source would constrain the cosmology \citep{zitrin18, 2021A&A...645A..44W}, although this rate is comparable to that from the mass assembly of the lens.  

\begin{figure}
\centering
\epsscale{0.8}
\plotone{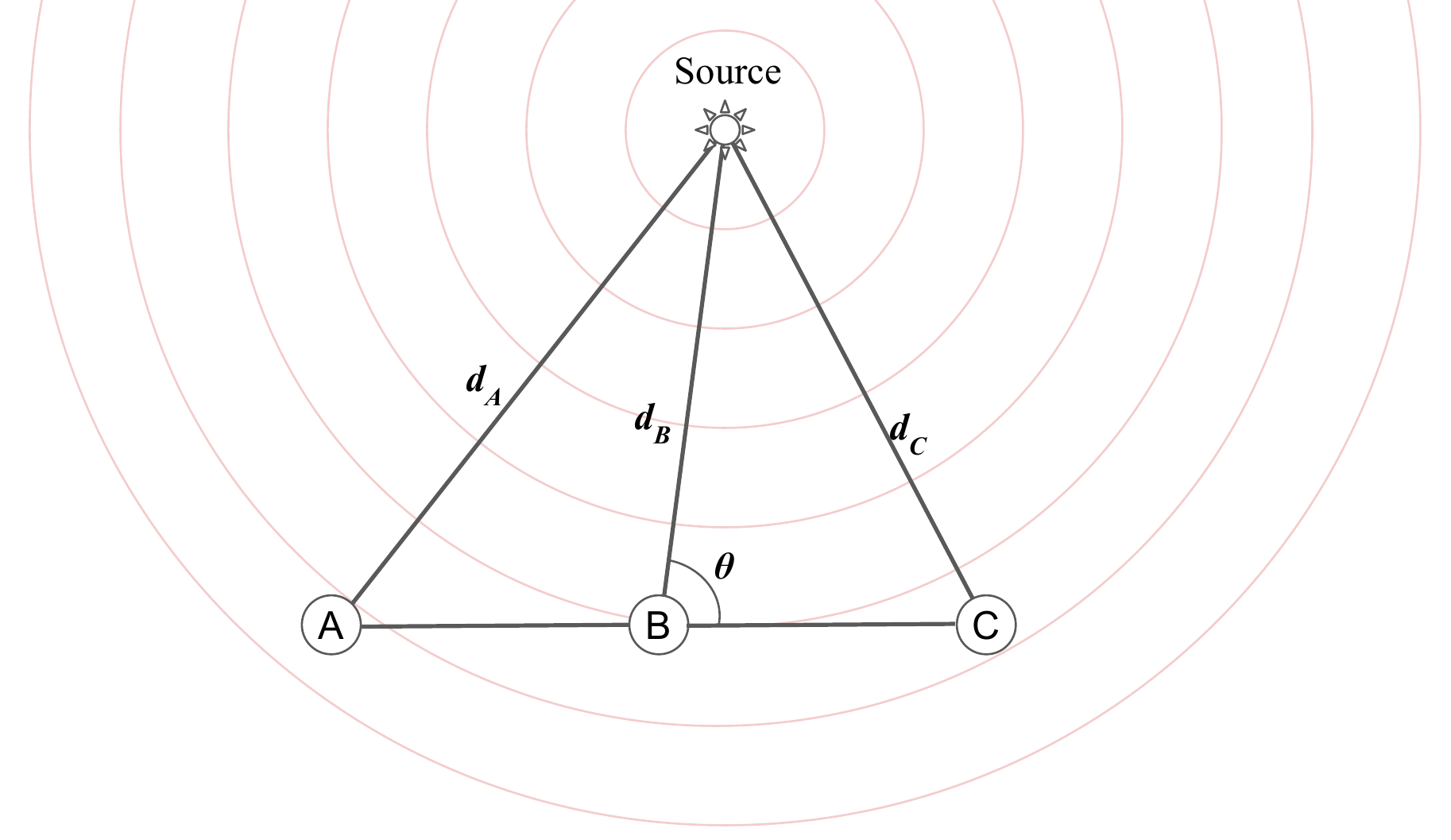}
\caption{Example of a detector configuration that can be used to measure the distance to a source from the curvature of its wave front. The signal will arrive at detector B before it is seen at detectors A or C. By comparing the arrival times at the three detectors we can infer the distance to the source. Note that we can only measure the difference in arrival times, not the distances $d_i$ directly. With two detectors the distance to the source is degenerate with the angular position on the sky $\theta$. With three detectors in two dimensions, or four detectors in three dimensions, this degeneracy is broken and the distance to the source can be inferred.}
\label{fig:curvature_overview}
\end{figure}

In this article, we propose a new geometric approach to constrain cosmology with FRBs. The wave front of an FRB will have a small curvature when it reaches our Solar System. By measuring the arrival time of the same FRB at four separate radiometers, this curvature can be measured directly and used to infer the distance.  (A similar idea has been applied to pulsar timing arrays in \citealt{2021PhRvD.104f3015D} and \citealt{2022MNRAS.517.1242M}.) An illustration of this idea is shown in Figure~\ref{fig:curvature_overview}. We will show that a constellation of detectors separated by 100 AU and observing at $\gtrsim 3$~GHz would be able to measure the cosmological distance to an FRB at 100 Mpc with an accuracy of $\sim$0.1\%. Sub-percent distances translate to sub-percent constraints on cosmological parameters via the redshift--distance relation once sufficiently into the Hubble flow. This idea also has the appeal of being a geometric measurement of distance in cosmology akin to parallax, not requiring the assumptions of other distance methods (e.g. the standardization of Type Ia supernovae; \citealt{1993ApJ...413L.105P}; or the $\Lambda$CDM model for cosmological density fluctuations; \citealt{2003moco.book.....D}).

The proposed experiment requires measuring the arrival times of FRBs with sub-nanosecond precision for Solar System-sized baselines. Sub-nanosecond timing is regularly done with very long baseline interferometry (VLBI; \citealt{SCHUH201268}).  With Earth-scale baselines, VLBI timing has allowed measuring the angular coordinates of thousands of radio sources with milli-arcsecond errors (\citealt{2002ApJS..141...13B}; and $\sim10\,\mu$-arcsecond for Sagittarius A$^*$ and the black hole at the center of M87; \citealt{2019ApJ...875L...1E}).
VLBI has also been done in space, with the 10\,m dish on the Spektr-R satellite as part of the RadioAstron program 
successfully finding correlations with a 30 Earth radius baseline \citep{1998Sci...281.1825H, 2013ARep...57..153K}. \tmp{VLBI is now regularly being done on FRBs \citep{Marcote2017, 2020Natur.577..190M, kirstenVLBI}.}

We also require the detectors to be localized with an accuracy of several centimeters over Solar System scales. Modern positioning systems can localize global navigation satellite systems (GNSS; such as the United State's GPS, Russia's GLONASS, China's BeiDou and Europe's Galileo) to meter or even centimeter precision \citep[e.g.][]{misraenge}. Millimeter accuracies on distance measurements are obtained with optimizations (longer integrations, having well-localized places on Earth for a differential measurement) that are able to break the integer wavelength ambiguity in the carrier phase.  Using very similar methods to GNSS systems, precision ranging is regularly performed to NASA satellites across the Solar System by the Deep Space Network,\footnote{\url{https://www.nasa.gov/directorates/heo/scan/services/networks/deep_space_network/}} reaching several meter precision in the separation and with the potential for improvement \citep{2009IPNPR.177C...1N}.  The Deep Space Network has executed precise radio ranging to the Pioneer, Voyager and New Horizons missions to the outer Solar System.

While placing radio telescopes on different sides of the Solar System would require a significant effort, a further motivation is that cosmology is near the end of how well we can measure parameters with traditional techniques.  Much of the useful cosmic volume for constraining dark energy will soon be mapped, and known techniques for constraining cosmological parameters cannot achieve errors below $\sim 1$\% on most parameters.  Currently there are 10\% tensions in the Hubble parameter determined with the most accepted techniques, namely the cosmic microwave background and Type Ia supernovae calibrated with parallaxes and then Cepheid variable stars \citep{Riess_2019, 2021CQGra..38o3001D}. If future measurements indicate a few-percent deviation from the $\Lambda$CDM expectation, would the community believe it?  The proposed experiment offers a potential way forward, enabling geometric sub-percent distance measurements where the major sources of error appear to be under the instrumentalist's control. We show that a \emph{single} FRB at $d=200\,$Mpc could be used to make a 1\% measurement of the Hubble constant for radio telescopes at Solar radii of $\gtrsim 20~$AU, with the 1\% set by the uncertainty in the peculiar velocity of the host galaxy. This measurement is comparable to the SHOES 1\% measurement of the Hubble constant using $\approx 100$ Type Ia supernovae with a median distance of $d=200\,$Mpc \citep{2021arXiv211204510R}. Each supernova provides a $10\%$ distance measurement, and so $\approx 100$ supernovae are required for 1\% precision; similar averaging can be done in our proposed experiment to achieve sub-percent measurements of the Hubble constant.  We further show that $\sim 100~$AU baselines could achieve 1\% constraints on a single FRB out to $d = 3000\,$Mpc or $z= 1$.

Perhaps the most apt comparison is with gravitational wave standard sirens.  Forecasts for a proposed next-generation instrument, the Einstein Telescope, are $\sim 1$\% distance errors to a low redshift and $\sim 5\%$ percent to $z=1$ \citep{2020JCAP...03..050M}, with these predictions reliant on assumptions about source demographics and the identification of electromagnetic counterparts. Our proposal offers the potential for higher precision even from a single FRB.

There are several applications beyond the obvious Hubble constant and dark energy science. Another cosmological parameter for which it would be interesting to push beyond current constraints is the spatial curvature.  If slow-roll inflation starts with a comoving Hubble radius that is comparable to our present-day horizon, the Universe could be more curved than the part in $10^5$ inflationary expectation, and there might be anthropic reasons for why we inhabit a region with an enhanced negative spatial curvature \citep{2006JHEP...03..039F}.  Additionally, the proposed extremely long baseline instrument could potentially measure the mass distribution in the outer Solar System, the distance to Galactic pulsars, constrain clumpy dark matter models, resolve the radio emission from pulsars, more directly measure the density fluctuation spectrum of ISM turbulence, and reach interesting sensitivities to $< 100\, \mu $Hz gravitational waves.

 This paper is organized as follows.  We present the idea for distance constraints from wave front curvature in \S~\ref{sec:geometric_dist}, and we argue that FRBs are likely the only extragalactic source that is sufficiently point-like for this method in \S~\ref{sec:sources}.  We then show that various systematics, namely ISM scattering, dispersion, and gravitational time delays, appear to be under control for $\nu \gtrsim 3\;$GHz, with gravitational time delays limiting the precision for $\gtrsim 100\;$AU antenna separations (\S~\ref{sec:systematics}; but also fully removable with two more detectors). We discuss the feasibility of calibrating the antennas' space-time locations to within one wavelength in \S~\ref{sec:calibration}.  Finally, \S~\ref{sec:discussion} synthesizes our estimates for the timing noise to forecast the achievable cosmic distance constraints, and \S~\ref{sec:otherscience} highlights other sciences applications that could be addressed with Solar System-scale interferometry. Appendix~\ref{sec:curvedspace} generalizes our calculations to curved space, and the other appendixes add details to some of the systematics calculations.


\section{Geometric distance measurements with time delays}
\label{sec:geometric_dist}

Here we describe how direct cosmological distances can be ascertained using the time delays between radiometers. 
This section presents a Euclidean-space derivation, which gives the essential idea and is a good approximation for sources that are not at appreciable redshifts. We generalize this derivation to Friedman-Robertson-Walker (FRW) space-times in Appendix~\ref{sec:curvedspace}.  In  flat FRW space-times, the distances in the Euclidean derivation are simply mapped to the comoving light-travel distances such that the following carries over directly.

\begin{figure}
\centering
\epsscale{0.6}
\plotone{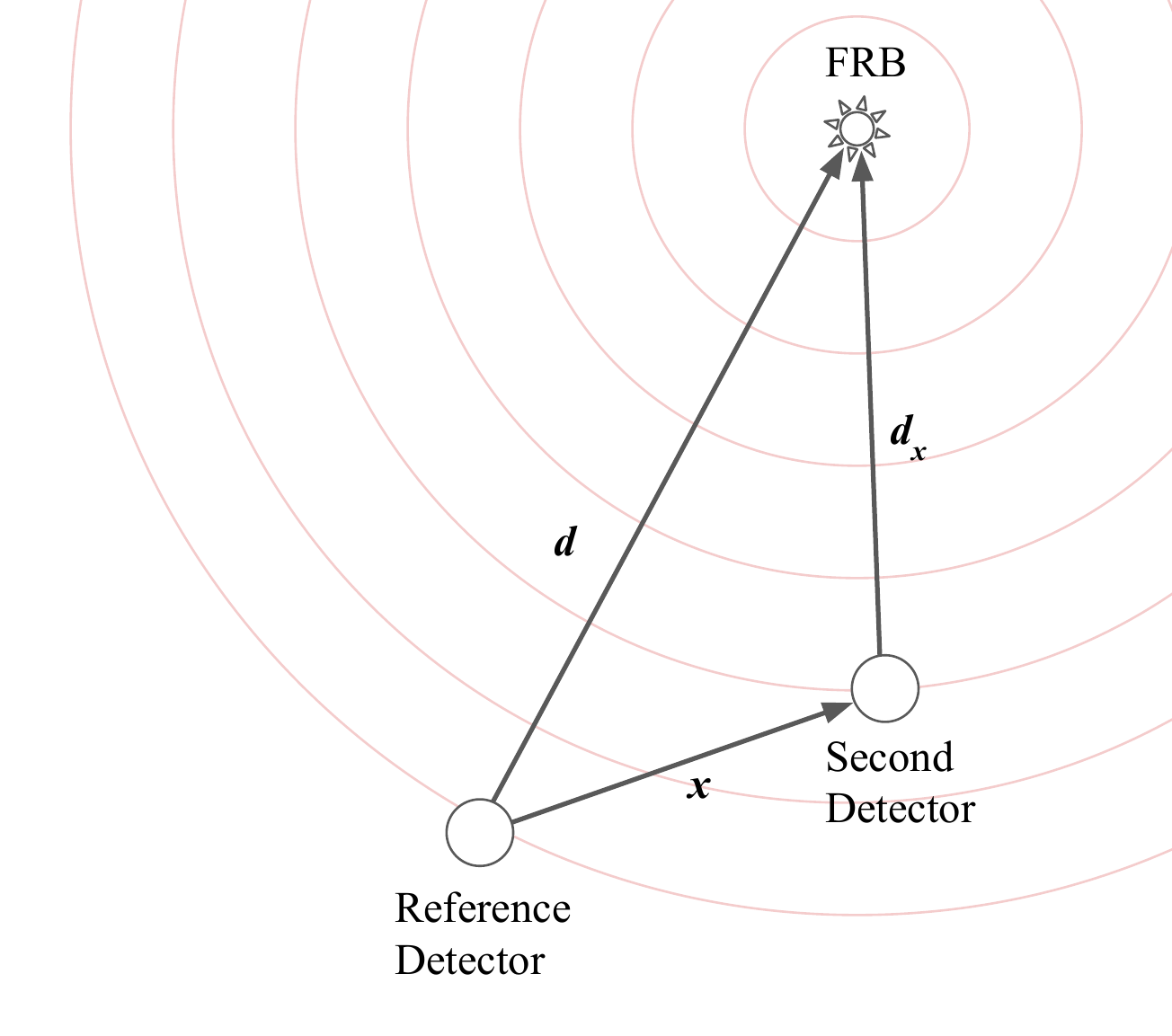}
\caption{The triangular configuration of two detectors and the source used for timing.}
\label{fig:detector_geometry}
\end{figure}

Let us first consider the case of two radiometers. Without loss of generality, we label one of our detectors the ``reference detector,'' and we define a coordinate system
so that this detector is at the origin. We label the position of the source as $\bfd$, the position of a second
detector as $\bfx$, and the vector between the second detector and the source as $\bfd_x$. Symbols in bold characters
refer to vectors, while symbols in regular fonts refer to the lengths of the corresponding vectors.
An illustration of this configuration is shown in Figure~\ref{fig:detector_geometry}.

The distance between the second detector and the source can be written as
\begin{align}
    d_x &= |\bfd - \bfx|= \sqrt{d^2 + x^2 - 2 \bfx \cdot \bfd},
    \label{eqn:lawofcosines}
\end{align}
Since $x \ll d$ (i.e. the distance between the two detectors is much
smaller than the distance to the source), we expand the previous equation in $x / d$, yielding
\begin{equation}
    d_x = d - \bfx \cdot \hatd + \frac{1}{2 d} \left(x^2 - (\bfx \cdot \hatd)^2\right) + \mathcal{O}(x^3/d^2),
\end{equation}
where $\hatd$ is the unit vector in the direction of $\bfd$.


The arrival time of the signal at the second detector relative to the first one is then given by:
\begin{align}
  c  \Delta t &= d_x - d = -  \bfx \cdot \hatd + \underbrace{\frac{1}{2\,d} \left(x^2 - (\bfx \cdot \hatd)^2\right)}_{c \Delta t_d} + \mathcal{O}\left(\frac{x^3}{d^2} \right).
    \label{eq:time_delay}
\end{align}
The $- \bfx \cdot \hatd$ term is the difference in time for a
flat wave front to travel past each of the detectors. The second term in the equation labeled $\Delta t_d$ is
the ``distance-dependent time delay'' and is quadratic in $x$ due to the fact that the wave front is not flat
when the source is at a finite distance.

Assuming that the detector geometry is known to better than the timing precision (the details of which will be discussed in Section~$\ref{sec:calibration}$), this equation has three unknown variables: two for the angular position of the source on the sky and one for the distance to the source. With at least four detectors measuring three independent time delays this system is fully constrained and can be used to infer the distance to the source.\footnote{Another way of doing the constraint counting is to note that four detectors yield the four equations: $|\bfd - \bfx_i|^2 = c^2(t_{\rm FRB} - t_i)^2$, where $i$ indexes each detector.  The time of the FRB $t_{\rm FRB}$ is one unknown and so this leaves just enough constraints to measure $\bfd$, assuming that the space-time positions of the detectors are known. This is not unlike the four satellites for precise GNSS positioning; there the fourth unknown is the signal arrival time at the receiver, as in the typical setup the receiver does not have a high precision clock.} The exact accuracy of this measurement depends on the full detector configuration. For typical configurations, we can approximate the fractional distance uncertainty $\sigma_d / d$ as
\begin{align}
    \frac{\sigma_d}{d} \sim \frac{\sigma_t}{\Delta t_d}
\end{align}
where $\sigma_t$ is the accuracy of a time delay measurement on a given detector pair. This expression was confirmed experimentally using Fisher matrix calculations as well as a Markov Chain Monte Carlo (MCMC) implementation in the \textsf{emcee} package \citep{foremanmackey13}. \tmp{For a range of different four-detector configurations where the detectors are all separated by similar projected baselines, we find that the error is approximately $\Delta t_d$. The errors quickly become smaller for more than four detectors owing to the large number of detector pairs, scaling roughly as $ \propto 1/\sqrt{(N_A - n_p)(N_A - n_p-1)/2}$ where $N_A$ is the number of detectors and $n_p$ is the number of additional parameters, being $n_p=2$ for the minimal scenario considered in this section where the additional parameters are the source angular coordinates  -- the scenario that we find applies until the baselines surpass 100~AU (\S~\ref{sec:discussion}).}

The distance-dependent time delay $\Delta t_d$ can be rewritten as
\begin{equation}
\Delta t_d = \frac{x_\perp^2}{2\, c\, d } = 120 ~\text{ns}~ \left(\frac{x_\perp}{100~\text{AU}}\right)^2 \left(\frac{d}{100~\text{Mpc}} \right)^{-1},
\end{equation}
where $1\;$AU is the Earth-Sun distance and $$\bfx_\perp \equiv \bfx - (\bfx \cdot \hatd)\hatd$$ is the projection of the detector separation $\bfx$ that is perpendicular to the line of sight. 

This time delay is small, but in principle can be measured by correlating the electromagnetic waves at each detector, using the detectors as an extremely large baseline for VLBI. As will be shown in Section~\ref{sec:timing}, a couple factors conspire so that for wide bandwidth measurements the time delays likely can be measured with an uncertainty of $\sigma_t \sim \nu^{-1}$ for $\nu \gtrsim 3\,$GHz. The fractional distance uncertainty of a single measurement can then be approximated as
\begin{align}
  \frac{\sigma_d}{d}
  \sim \frac{\sigma_t}{\Delta t_d}
  = 0.17 \%
    \left( \frac{x_\perp}{100~\text{AU}} \right)^{-2}
    \left( \frac{d}{100~\text{Mpc}} \right)
    \left( \frac{5~\text{GHz}}{\nu} \right) \left( \frac{\sigma_t }{\nu^{-1}} \right).
\end{align}
Thus, the sensitivity scales with the square of the baseline, and baselines of at least several AU are required to make cosmologically interesting distance measurements at any frequency where VLBI has been performed. A system with a baseline of 100~AU observing at 5~GHz could obtain sub-percent geometric distance measurements to individual sources out to several hundred megaparsecs. 

The curvature wave front effect is related to the more astronomically familiar parallax effect, as the extra delay from curvature can be thought of as different baselines detecting the source at slightly different locations on the sky.  Indeed, if we have three equally spaced detectors each in a line with a spacing of $x_\perp$, the parallax angle between the centers of the left and right baseline is $x_\perp/d$ and each pair of adjacent antennas has an angular resolution of $\sigma_\theta \sim c \,\sigma_t/x_\perp $.  Therefore $d$ can be measured to a fractional precision of $\sigma_\theta/(x_\perp/d) \sim \sigma_t/\Delta t_d$, the same relation we found from timing considerations.  In the language of parallax, the reason four detectors are needed is that each baseline localizes the position of a source to a line with a girth of $\sigma_\theta$ (i.e. the angular position is unconstrained perpendicular to the baseline).  Thus, only when the three detectors are in perfect alignment is this sufficient to measure the parallax, but alignment is no longer required once a fourth detector is added.

\section{FRBs as the best source candidate}
\label{sec:sources}

Interferometric time delay measurements require the angular size of the source to be smaller than (or at least comparable to) the angular resolution of the baselines for the two antennas to receive identical (or nearly so) waveforms.  This means that the source must be smaller than $\theta = \lambda/x_\perp =  8\times 10^{-4}~({\rm 100 AU}/x_\perp) (\nu/5~\text{GHz})^{-1}~\mu$-arcsecond, where $\lambda = c/\nu$ is the wavelength, or equivalently that the physical size of the source must satisfy\footnote{Technically the $d$ that appears here is the angular diameter distance, unlike in all other sections, where (in flat space) it is the comoving light travel distance.}
\begin{eqnarray}
\ell  \lesssim \frac{\lambda d}{x_\perp} =  1\times 10^{12} ~\left (\frac{\rm 100~ AU}{x_\perp} \right) ~\left( \frac{\nu}{ 5 \, {\rm GHz}}\right)^{-1}~\left(\frac{d}{100 ~\text{Mpc}} \right)~\text{cm}.
\end{eqnarray}
  We can rewrite this equation in terms of the fractional distance and timing errors:
\begin{eqnarray}
\ell  &\lesssim& \frac{1}{2} (\sigma_d/d )( \nu \, \sigma_t)^{-1} x_\perp, \label{eqn:ellxperp}\\ 
 &\lesssim& \sqrt{\frac{c \, d \;\sigma_d/d  }{2\, \nu^2 \,\sigma_t}}  = 10^{12.5} \left(\frac{d}{100 \,\text{Mpc}}\right)^{1/2} \left(\frac{\sigma_d/d}{0.01}\right)^{1/2} \left(\frac{\sigma_t}{\nu^{-1}}\right)^{-1/2} \left(\frac{\nu}{5\, \text{GHz}}\right)^{-1/2} {\rm cm},
 \label{eqn:ellprec}
\end{eqnarray}
where we used that $\sigma_d/d = \sigma_t/\Delta t_d \sim 2\, c\,d\, \sigma_t/x_\perp^2 $ and have grouped terms in a manner that reflects the timing noise $\sigma_t$ is likely to be of the order of $\nu^{-1}$ (\S~\ref{sec:systematics}).  Equation~\ref{eqn:ellxperp} suggests that, for a timing precision of $\sigma_t \sim \nu^{-1}$, the source needs to be $\gtrsim 100\times$ smaller than the baseline separation for percent-level cosmology.

One potential source candidate in the sub-millimeter is the shadows of supermassive black holes \citep{2000ApJ...528L..13F, 2019ApJ...875L...1E}.  Supermassive black hole shadows have sizes of $\sim 3GM/c^2 = 4\times10^{14} (M_{\rm BH}/10^9 M_\odot)~$cm. Thus, to be unresolved in the sub-millimeter band at $\nu \sim 200~$GHz, a black hole mass of $\lesssim 10^6 ({d}/{100 \,\text{Mpc}})^{1/2} M_\odot$ is required for 1\% distance constraints if timing can be performed to $\sigma_t \sim \nu^{-1}$.  It is unlikely there are bright enough candidates at these masses: This emission is relatively faint as a visible shadow requires the accretion disk to be in a radiatively inefficient state, with the $6\times10^9~M_\odot$ M87 black hole at just 16 Mpc having a flux density of $\sim 0.5\,$Jy in the submillimeter and with this black hole being more luminous than other similarly distant active galactic nuclei with compact radio emission \citep{10.1093/mnras/stw391}.  M87 is accreting at $\sim 10^{-5}$ its Eddington limit, and a radiatively inefficient state may still be possible even for $10^{-2}$, with the luminosity scaling quadratically with the mass accretion rate \citep{2008NewAR..51..733N}, such that it is not impossible that such a favorable class of black hole shadows exist. 

Another source candidate is the faint radio afterglows of gamma ray bursts (GRBs). The afterglow emission region has a size of $\ell \sim c \, t \, \theta_J = 3\times10^{13} ~(t/{\rm 1 ~day}) (\theta_J/0.01)~$cm,
where measurements indicate that the smallest jet opening angles have $\theta_J \sim 0.01$  \citep{2004RvMP...76.1143P}.  Therefore, even when observed a day after the GRB, which is unusually early \citep{2004RvMP...76.1143P}, GRB radio afterglows are somewhat too large for this experiment.  

Another possibility is to abandon the radio and do interferometry in the optical, where distance estimates with precision $\sigma_d/d = 0.01$ for a source at $d=100\,$Mpc requires a baseline of \emph{just} $0.1~$AU if timing can be performed to $\sigma_t = \nu^{-1}$.  Now, looking past the impediment that direct interferometry of the optical light from the two paths is likely required, making achieving such baselines difficult to say the least, the most obvious optical point source -- quasars' accretion disks -- are still too large for the compactness requirement of $\ell \lesssim 10^{10}$cm in the optical band, with inferred sizes of thousands of AU \citep{2010ApJ...712.1129M}.

This brings us to FRBs, our best source candidate. Typical FRB durations of milliseconds put an upper bound on the variability timescale, $\delta t_{\rm var}$, which suggests that the underlying source is $\lesssim c \ \delta t_{\rm var}$ or $\lesssim  1000$\;km, $4-5$ orders of magnitude smaller than the size requirement at gigahertz frequencies for baselines capable of precision cosmology.  Some FRBs even show variability on microsecond or even nanosecond timescales \citep{2021NatAs...5..594N, 2022NatAs...6..393N}.  
There is substantial evidence that FRB emission originates from magnetars (which are highly magnetized neutron stars; \citealt{petroff19}), including the detection of an FRB from a Galactic magnetar \citep{magnetar, chimemagnetar}. In one class of magnetar models, the FRB originates from near the $\sim 10~$km surface of the neutron star \citep[e.g.][]{2020MNRAS.494.2385K}.  However, in the alternative class of magnetar emission models, the FRB originates from a relativistic shock further out \citep[e.g.][]{Metzger:2019una}.  In this model, the observed emission region should be $\ell <  (2 \Gamma^2 \, c \, \delta t_{\rm var})/\Gamma$, i.e. larger by a factor of the Lorentz factor $\Gamma$ relative to $c \, \delta t_{\rm var}$, but still smaller than we require for $\Gamma \lesssim 10^{4.5}$ if we take $\delta t_{\rm var} = 1$~ms, which is well above what is allowed for some FRBs, and the terms in parentheses in Equation~\ref{eqn:ellprec} evaluate to unity (furthermore the shock would be at an immense radius of $\gtrsim 10^{17}$cm).  

Multi-path diffractive propagation from electron inhomogeneities in the FRB host galaxy can make the effective image size larger.\footnote{\tmp{This image broadening also happens from plasma interactions within the Milky Way, which we discuss in \S~\ref{sec:scattering}.}}  This effect is called ``scattering.''  The scattered image size is related to the scattering timescale by $c \tau_{\rm sc} \sim \theta_{\rm screen}^2 d_{\rm screen}$, where $\theta_{\rm screen}$ is the angle from the FRB to the illuminated part of the scattering screen at a distance $d_{\rm screen}$.  Then the effective size of the FRB emission region is $\ell =  \theta_{\rm screen} d_{\rm screen} = \sqrt{c \tau_{\rm sc} d_{\rm screen}}  = 1\times 10^{11} (d_{\rm screen}/0.1\; {\rm pc})^{1/2} (\tau_{\rm sc}(1~\GHz)/1\; {\rm ms})^{1/2} (\nu/5 \,{\rm GHz})^{-2}$~cm, assuming that the scattering time scales in the standard way as $\nu^{-4}$ (\S~\ref{sec:scattering}).  A $\tau_{\rm sc}(1\ \text{GHz}) \sim 1 $\,ms scattering timescale contributed by the host galaxy is detected in a significant fraction of FRBs \citep{thornton13, qiu20}, although for some $\tau_{\rm sc}(1\ \text{GHz})$ is constrained to be $\lesssim 1\,\mu$s at $\nu=1~$GHz \citep{2021NatAs...5..594N, Cho_2020}.  The choice $d_{\rm screen}=0.1\;$pc is motivated by the radius of the magnetized nebula hundreds of years after the supernovae, a timescale which explains the persistent radio emission observed around some repeating FRBs \citep{10.1093/mnras/sty2417}.  \tmp{Indeed \citet{Marcote2017} constrained with VLBI the persistent radio emission to be $<0.7~$pc for FRB~121102.}  In conclusion, unless the millisecond scattering occurs at $d_{\rm screen} \gtrsim 60 (\nu/5 \;{\rm GHz})^{3}$~pc , the scattered image of an FRB at $d=100~$Mpc is likely sufficiently small.

\section{Timing measurement accuracy}
\label{sec:systematics} \label{sec:timing}


The proposed experiment correlates the electromagnetic
waves at each detector, using the detectors as an extremely large baseline for VLBI. \tmp{Often for VLBI astrometry, timing is done using the group delay, which is the rate of change of phase over the band (not requiring the absolute phase).  This leads to a timing error of}
\begin{align}
    \sigma_{t, \rm group}^{\rm VLBI} &= \frac{1}{2 \pi\Delta \nu_{\rm RMS}~\text{SNR}}, \\
             &= 0.016~\textrm{ns}~\left(\frac{\Delta \nu_{\rm RMS}}{1~\textrm{GHz}}\right)^{-1} \left(\frac{\mathrm{SNR}}{10}\right)^{-1},
             \label{eqn:sigmat}
\end{align}
where SNR is the signal-to-noise ratio at which correlations between two detectors are detected and
$\Delta \nu_{\rm RMS}^2 \equiv \text{SNR}^{-2} \{\int d\nu \, \nu^2 \, d({\rm SNR}^2)/d\nu -  [\int d\nu \, \nu \, d({\rm SNR}^2)/d\nu]^2\}$ is the effective RMS bandwidth squared \citep{1970RaSc....5.1239R, 2017isra.book.....T}.\footnote{If correlating two instruments with different flux sensitivities, the SNR that appears in Eqn.~\ref{eqn:sigmat} is the geometric mean of the SNR of each instrument to detect the flux.}  For uniform signal to noise over a continuous bandwidth of $\Delta \nu$ then $\Delta \nu_{\rm RMS}  = \Delta \nu/(2\sqrt{3})$.  \tmp{While the group delay is less prone to systematics (namely clock phases and the Earth's atmosphere), using the absolute phase improves the timing noise over $\sigma_{t, \rm group}^{\rm VLBI}$ by approximately the factor $\Delta \nu_{\rm RMS}/\nu$.  One gains from using the phase delay once the space-time detector positions are known more precisely than the group delays, although sufficient bandwidth is needed to discriminate the correct phase fringe ($\nu/\Delta \nu \ll {\rm SNR}$). See Appendix~\ref{ap:fittingdisp} for more discussion and we return to this timing when we discuss dispersion in \S~\ref{sec:dispersion}.}  


Other considerations also affect the timing precision.  Namely, inhomogeneities in the intervening plasma that alter the path traveled by light (scattering; \S\ref{sec:scattering}) and ones that result in a relative difference in the average group velocity between the detectors' sightlines (differential dispersion; \S~\ref{sec:dispersion}). Additionally, gravitational time delays turn out to be important for baselines of $x\gtrsim 100\,$AU (\S~\ref{ss:gravitationaldelays}).



\subsection{Scattering}
\label{sec:scattering}

Scattering refers to the geometric time delay and smearing of the arrived waveform from density inhomogeneities in the intervening plasma. The Milky Way's interstellar medium (ISM) can lead to different delays and smearing along the sightline to each detector.  Observations of pulsars and some FRBs above the Galactic disk suggest a scattering time of
\begin{equation}
    \sigma_t^{\rm sc} \sim 30~\textrm{ns}~\left(\frac{\nu}{1~\textrm{GHz}}\right)^{-4} \sin(b)^{-2.5}
    \label{eqn:sc}
\end{equation}
for Galactic latitude ($b$) sightlines above the disk, with sightlines having a factor of several scatter around this `average' \cite[the coefficient in Equation~\ref{eqn:sc} is the NE2001 model value given in the first reference]{2002astro.ph..7156C, cordes16, 2021NatAs...5..594N, 2022NatAs...6..393N}.  
  \tmp{The frequency scaling of $\sigma_t^{\rm sc}$ is important as most measurements of scattering are at smaller frequencies than are ideal for the proposed experiment.
  We adopt a $\nu^{-4}$ scaling in Equation~\ref{eqn:sc}, which is used in our later forecasts.
A slightly steeper scaling of $\nu^{-4.4}$ is the expectation from Kolmogorov turbulence \citep{1985MNRAS.213..591B}. In models where the scattering owes to the refractive lensing of a small number of current sheets \citep[which have been developed to explain pulsar scintillation data;][]{2014MNRAS.442.3338P}, the average scaling would be $\nu^{-4}$ and with system-to-system scatter about this scaling}. Measurements are on-average consistent with $\nu^{-4}$ and show significant system-to-system scatter \citep{ 2004A&A...425..569L, 2019ApJ...878..130K, 2021MNRAS.504.1115O}.
 
For the baselines and frequencies we are considering, the structures that contribute to the scattering at each antenna are approximately uncorrelated, as this scattering time corresponds to light taking paths that differ by a transverse length of
\begin{align}
    R_T &= \sqrt{2 \, L \, c \,\sigma_t^{\rm sc}} \sim 0.006~\text{AU} ~\sin(b)^{-1.3} \left(\frac{\nu}{5~\text{GHz}}\right)^{-2} \left( \frac{L}{1~\text{kpc}} \right)^{1/2}.
    \label{eqn:RT}
\end{align}
 Since $R_T$ is much less than our $x\gg1$~AU baseline lengths for relevant frequencies, the scattering seen by each detector will be approximately uncorrelated. 
 
 When the scattering is strong ($\sigma_t^{\rm sc} \gtrsim \nu^{-1}$), the light travels over multiple paths with significant phase differences resulting in the paths interfering at the antennas. This interference smears out the electric field correlations, making them harder to detect \citep{narayan}.  Thus, it becomes increasingly difficult with decreasing frequency to measure the time delays between different detectors. When $\sigma_t^{\rm sc} \lesssim \nu^{-1}$ the different paths add constructively and the scattering becomes weak. Using Equation~\ref{eqn:sc}, this corresponds to $\nu\gtrsim 3 \sin(b)^{-0.8}~$GHz. Weak scattering has the effect of adding an additional delay to each sightline but not smearing out temporally the correlations. For a given system, the frequency scaling of $\sigma_t^{\rm sc}$ should be the same across the weak and strong limits \tmp{when there are enough structures so that the phase field is Gaussian random \citep{1989MNRAS.238..995G} and, furthermore, weak refractive scattering always scales as $\nu^{-4}$}. 
 
 In conclusion, Milky Way ISM scattering favors targeting $\nu \gtrsim 3\,$GHz, as at lower frequencies scattering will exceed our $\nu^{-1}$ timing-noise goal and additionally its interfering nature will make $E$-field correlations more challenging to detect.

\subsection{Differential dispersion} \label{sec:dispersion}

Dispersion in interstellar plasma will introduce a frequency-dependent time delay owing to the wave front traveling with a group velocity that is less than the speed of light.  This delay is equal to 
$\tau_{d} = \kappa \int_0^d n_e \,dx_\parallel$, where $\kappa={e^2}/(2 \pi \, m_e \, c \,\nu^2)$.
%
%
Here $n_e$ is the electron density and this integral is performed along the entire line of sight to the source. For observations at 5\;GHz, and a typical $\int_0^d n_e \, dx_\parallel$ (called the `dispersion measure') in current FRB samples of $500~\textrm{pc}~\textrm{cm}^{-3}$, the corresponding time delay is 100~ms. \tmp{
For this experiment the overall $\tau_d$ does not affect timing, but differences in the $\tau_d$ between the radiometer sightlines do.  If large enough -- which we show is likely --, these delays need to be fit for and removed. (The higher-order effect from deviations from straight-line paths is included in the scattering delay considered in \S~\ref{sec:scattering}.)}

\tmp{The largest contribution to  $\tau_d$ differences to each detector is likely turbulence in the ionized ISM.  Appendix \ref{sec:dispturb} considers this case for a power spectrum parameterized as $P_e(k) = (2\pi)^3 C_n^2\, k^{-11/3}$ and finds that the contribution that looks like wave front curvature is
\begin{align}
    \sigma_{t}^{\rm disp} \approx 90~\textrm{ns}
    \left(\frac{L}{0.1~\textrm{kpc}}\right)^{1/2}
    \left(\frac{x_\perp}{100~\textrm{AU}}\right)^{5/6}
    \left(\frac{\nu}{5~\textrm{GHz}}\right)^{-2} \left( \frac{C_n^2}{5 \times 10^{-17} \textrm{cm}^{-20/3}}\right)^{1/2}.
    \label{eqn:sigmadisp}
\end{align}
 In the Solar neighborhood measurements find $C_n^2 \sim 5 \times 10^{-17} \textrm{cm}^{-20/3}$ \citep{1981Natur.291..561A, draine11}. We have chosen the $L$ so that when the parentheses evaluate to unity the scattering measure SM$\equiv C_n^2 L =5 \times 10^{-18}\text{kpc~cm}^{-20/3}= 1\times 10^{-4}~$kpc~m$^{-20/3}$ aligns with high Galactic latitude values in the NE2001 model \citep{2002astro.ph..7156C}.  Models where the scattering does not owe to turbulence but instead AU-scale current sheets (which better describe pulsar scintillation data) would result in smaller $\sigma_{t}^{\rm disp}$ owing to the bluer $P_e$ \citep{2006ApJ...640L.159G, 2014MNRAS.442.3338P}.
}

In addition, more local gaseous structures can contribute to the dispersive delays. The differential electron column from the heliosphere at $R_{\rm HS}\approx 100\,$AU can give rise to a time delay just like ISM density fluctuations.  Outside the heliosphere, the density rises by almost a factor of one hundred to $n_{\rm IS} = 0.1\,$cm$^{-3}$ \citep{2013Sci...341.1489G}.  The differences in electron column from these positional differences result in timing differences of
\begin{equation}
  \sigma_{t}^{\rm disp}  \sim  \kappa \frac{n_{\rm IS} x_\perp^2}{R_{\rm HS}} = 0.5 ~\text{ns}~~ \left(\frac{\nu}{5~\text{GHz}} \right)^{-2} \, \left(\frac{x_\perp}{100~\text{AU}}\right)^2.
\label{eqn:sigmadisphelio}
\end{equation}
While this component scales similarly with $x_\perp$ as $\Delta t_d$, it is also small.  Additionally, if a terrestrial antenna is used in the proposed constellation, the Solar wind leads to several nanosecond dispersive delays at $5\,$GHz \citep{2009IPNPR.177C...1N}.

 There are also timing differences from cosmological structures such as circumgalactic clouds, but we find that these are negligible owing to their large extents.
 
\tmp{ Pulsar timing arrays empirically constrain dispersion variations.  Most pulsars followed by the NANOGrav timing array show variations in their dispersion measure of $\sim 100 ~{\rm ns} ~({5~\text{GHz}}/\nu)^{2}$ over $\sim 10\,$yr of monitoring, which probes electron variations on lengths of $\sim 100~$AU \citep{Jones2017}.  For most, the dispersion changes may be dominated by the pulsar motion in its overdense vicinity, but the variation is surprisingly close to our estimate from ISM turbulence.}
 \\

\tmp{Thus, the differential dispersive delays are likely to be much larger than the $\nu^{-1}$ timing precision specification at the $\nu \lesssim 10~$GHz at which FRBs have been detected. Fortunately, dispersive delays can be removed by fitting simultaneously for them using their distinct $\nu^{-2}$ frequency dependence.   Fitting out differential dispersion requires a wide bandwidth to not incur too large of a cost to the wave front timing sensitivity.  In Appendix~\ref{ap:fittingdisp}, we find that the timing noise with marginalization over dispersion is given by $\sigma_{t, \rm group}^{\rm VLBI}/2$ in the case where the absolute phase is used (and by $10 \, \sigma_{t, \rm group}^{\rm VLBI}\, [0.2 \, \nu]/\Delta \nu$ in the less applicable case where only the group delay is used), assuming a continuous bandwidth of $\Delta \nu$ and uniform signal to noise over this bandwidth.  This phase delay timing error increases over the case without fitting for dispersion by a factor of $ \approx \nu/(2 \,\Delta \nu_{\rm RMS})$.  \emph{Importantly, the error bars from correcting the dispersive delays do not depend on the size of these delays.}  This independence is well known from GNSS applications, where not correcting for the relative dispersion from Earth's atmosphere can result in $20~$m errors.}

\subsection{Gravitational time delays}
\label{ss:gravitationaldelays}
There will be stars, Solar System bodies, and even large-scale structure along the line of sight to an FRB that will introduce
different gravitational time delays to each detector. 
For a mass at a distance $d_M$ for which the impact parameter is much less than both $ \{d_M, |\bfd - \bfd_M|\}$, the difference in
the Shapiro time delays between the detector at $\bfx$ and the one at origin is given by \citep{1964PhRvL..13..789S}:
\begin{align}
    \Delta t_\text{grav} &= \frac{2 G M}{c^3} \text{ln}\frac{(b + \delta_b)^2}{b^2},
    \label{eqn:shapiro}
\end{align}
where $M$ is the mass of the star, $\bfb$ is the impact parameter for the first path (perpendicular to $\bfd$), and 
\begin{equation}
\delta_b =   |\bfb + \bfx - (\bfx\cdot \widehat{\bfd}) \widehat{\bfd}| - b \approx \widehat{\bfb} \cdot \bfx + \frac{1}{2b} \left[ x^2 - (\bfx\cdot \widehat{\bfd})^2\right] - \frac{1 }{2 b} \left(\widehat{\bfb} \cdot \bfx \right)^2 +{\cal O} \left(\frac{x^3}{b^2} \right) 
\end{equation}
is the difference in the impact parameter for the second path, which this expression relates to $\bfx$. (Recall that $\bfx$ is our antenna separation vector.)  Hats denote unit vectors, and vectors are the same as indicated in Figure~\ref{fig:detector_geometry}.  The term $- (\bfx\cdot \widehat{\bfd}) \widehat{\bfd}$ projects out the line of sight component of $\bfx$.   Assuming that the difference in impact parameter is small relative
to the impact parameter itself, we can expand Equation~\ref{eqn:shapiro} in terms of $\delta_b / b$
to obtain
\begin{align}
    \Delta t_\text{grav} &= \frac{2 G M}{c^3} \left(\frac{2 \delta_b}{b} - \frac{\delta_b^2}{b^2}\right) + \mathcal{O}\left(\frac{\delta_b^3}{b^3}\right),\\
    &= \frac{4 G M}{c^3 b^2} \left( \bfb \cdot \bfx + \frac{1}{2} \left[ x^2 - (\bfx\cdot \widehat{\bfd})^2\right]  - \frac{\left(\bfb \cdot \bfx \right)^2}{b^2} +{\cal O} \left(\frac{x^3}{b} \right) \right), \\
    &= \frac{4 G M}{c^3 b^2} \left( b \,x_\perp  \cos(\theta) - \frac{1}{2} x_\perp^2 \cos(2 \theta) +{\cal O} \left(\frac{x^3}{b} \right) \right), \label{eqn:tgrav}
\end{align}
where $\theta$ is the angle between $\bfx_\perp$ and $\bfb$.  The last line used $$\left[ x^2 - (\bfx\cdot \widehat{\bfd})^2\right]  - 2\left( \bfx \cdot \widehat{\bfb} \right)^2 = x_\perp^2  - 2\left( \bfx_\perp \cdot \widehat{\bfb} \right)^2 = -x_\perp^2 \cos(2 \theta),$$  Equation~\ref{eqn:tgrav} organizes the expansion into a dipole, quadrupole, etc.

The linear-in-$x_\perp$ dipolar term accounts for the majority of the contribution to $\Delta t_\text{grav}$. 
However, the linear term is identical to changing the direction of the FRB, which had resulted in $- \bfx \cdot \widehat{\bfd}/c$ in $\Delta t_d$ in Equation~\ref{eq:time_delay}.  Thus, the linear effect of the Shapiro time delay manifests as a small shift in the apparent direction of the source so that it appears in the direction $(\widehat{\bfd} - \alpha \, \widehat{\bfb})/\sqrt{1+\alpha^2}$, where $\alpha = 4 G M/(c^2b)$, which one might recognize as the deflection angle from lensing.  The effective $\alpha$ when accounting for the accumulation of all gravitational deflections is dominated by the large-scale structure of the Universe for sources at cosmological distances, resulting in arcminute-scale deflections for sources at high redshifts -- well known from studies of the cosmic microwave background \citep{2006PhR...429....1L}.\footnote{A worry is that the full linear-in-$x$ delay when adding the Shapiro effect to the geometric delay is $(\widehat{\bfd} - \alpha \, \widehat{\bfb})\cdot \bfx/c$, where $(\widehat{\bfd} - \alpha \, \widehat{\bfb})$ is not a unit vector, suggesting we might need an additional timing constraint to constrain the third component.  The non-unit vector term to leading order in $\alpha$ equals $\alpha^2 \widehat{\bfd}\cdot  \bfx/(2c) $, which is relatively large for cosmological values of the deflection angle $\alpha$.  However, the presence of this term is confusing as the linear Shapiro effect acts to tilt the wave front, which should be describable with a unit vector. The resolution is that the Shapiro effect does not account for the additional path length of the tilted wave front.  Accounting for this exactly cancels this term.}

The quadratic-in-$x_\perp$ quadrupolar Shapiro term is a source of noise for our distance measurement, with a value given by
\begin{align}
    \Delta t_\text{grav}^\text{quad}
           &= -0.025~\textrm{ns}   \cos(2 \theta)\left(\frac{x_\perp}{100~\textrm{AU}}\right)^2 \left(\frac{M}{M_\odot}\right) \left(\frac{b}{1~\textrm{ly}}\right)^{-2},
           \label{eqn:tgravquad}\\
            &=- 0.023~\textrm{ns} \cos(2 \theta)\left(\frac{x_\perp}{100~\textrm{AU}}\right)^2\left(\frac{M}{10^{13} M_\odot}\right) \left(\frac{b}{1~\textrm{Mpc}}\right)^{-2}.
           \label{eqn:tgravquadcosmo}
\end{align}
Let us concentrate first on the stellar case for which Equation~\ref{eqn:tgravquad} is expressed.
 The stellar surface density from the Solar Circle out of the Galactic disk is estimated to be  $2 \;M_\odot\,$Lyr$^{-2}$ \citep{2015ApJ...814...13M}. 
For the case of Poisson sampling of stellar locations from a spatially uniform probability density, the quadratic Shapiro term will be dominated by the closest stars to a sightline and so the parameters that yield $\Delta t_\text{grav}^\text{quad} = -0.025\,\cos(2 \theta)~$ns in Equation \ref{eqn:tgravquad} should be in line with the typical Shapiro delay from stars in the Galactic disk.\footnote{If there is a stellar structure that has a large quadrupole moment, the collective contribution of many stars would be important.  The most obvious structure is the Galactic disk; however, the Disk mostly contributes a dipole (a gradient) for high-latitude sightlines.}

Regarding dark matter halos and large-scale structure, most sightlines out to a gigaparsec will pass within the 300\,kpc virial radius of a $\gtrsim 10^{12}M_\odot$ halo \citep{mcquinn14}.  
Thus, this fact combined with Equation~\ref{eqn:tgravquadcosmo} suggests that the delay from dark matter halos is comparable to that of Disk stars.  However, a more detailed calculation in Appendix~\ref{ap:shapiro} shows that the dark matter halo contribution is generally larger (with an RMS of $0.1\;$ns for $d=100\;$Mpc and $x_\perp=100\;$AU) and increases roughly as the square root of the distance to the FRB.

The $\propto x_\perp^2$ dependence of the quadratic Shapiro delay means that, if this term is not removed, eventually increasing the baseline length does not increase the precision to which $\Delta t_d$ can be measured.  We find in \S~\ref{sec:discussion} that the Shapiro delay can become the dominant noise source at $x_\perp \gtrsim 100\,$AU.   As the quadratic Shapiro term is quadrupolar (c.f. Equation~\ref{eqn:tgravquadcosmo}), it is completely specified by two angular moments such that two additional detectors over the minimal four-detector constellation would enable isolating and subtracting it.\\


Now lets turn to near-field Shapiro delays from nearby Solar System bodies. For such bodies we cannot take the previous distant mass limit as $b\sim d_M \ll d$, where $d_M$ is the distance from the body to a detector.  We use primed values for these distances to a second detector, where again $b'\sim d_M'$.  Thus, we require a more complete expression for the Shapiro delay, given by  
\begin{equation}
    \Delta t_{\text{grav}, \text{SS}} = \frac{2 G M}{c^3} \left[ \ln \left(\frac{d_M + \sqrt{d_M^2 + b^2}}{{d'}_M + \sqrt{{d_M'}^2 + {b'}^2}}\right) - \frac{1}{2}\left(\frac{d_M}{\sqrt{d_M^2 + b^2}} - \frac{{d'}_M}{\sqrt{{d'}_M^2 + {b'}^2}}\right) \right].
    \label{eqn:shapironearby}
\end{equation}
In this nearby limit, the relative Shapiro time delay difference between two baselines is of order $2GM/c^3$.  In the extreme scenario in which all the mass in the outer Solar System were closer to one baseline than to another, the total effect would be of the order of $2G\,M/c^3 = 3\times10^{-2} (M/M_\earth)~$ns.  The Kuiper Belt is estimated to have a mass of $10^{-2}-0.1 \, M_\earth$ and a similar mass is for known trans-Neptunian objects \citep{2018AstL...44..554P}.  Thus, this amount of mass does not lead to problematic delays, and really the systematic will be smaller than the above estimate, likely reduced by the square root of the effective number of systems that contribute (which for the $\sim 100$ trans-Neptunian objects would be a factor of $\sim 10$ suppression).  

The hypothesized ``Planet 9,'' with a mass of $\sim 10 \,M_{\earth}$ and an orbit of $\sim 500~$AU \citep{2019PhR...805....1B}, would lead to timing noise at the $0.1~$ns level and could be a concern.  However, its mass would likely be measured and corrected by our proposed instrument, both by how its gravity deflects the orbits of the antennas  \citep[][and \S~\ref{sec:otherscience}]{2020arXiv200414192W} and by its effect on their onboard clocks (\S~\ref{sec:systematics}).


\section{FRB Rate}
\label{sec:FRBrate}

The clearest path is to directly target the known population of repeating FRBs, for which about fifty are currently known \citep[which has doubled since this article was released as a pre-print!]{2020ApJ...891L...6F, petroff21, 2023arXiv230108762T}.  The current sample of repeating FRBs with a known galactic host include FRB~20121102A (the original repeater) at a comoving light-travel distance of $d =800$~Mpc, FRB~20180916B at $d =150$~Mpc, FRB~20180301A and FRB~20190520B are somewhat further away with $d\approx 1000$~Mpc, FRB~20181030A is most likely at $20~$Mpc, and FRB~20201124A at $400$ Mpc  \citep{petroff21, 2021ApJ...919L..24B}.  Bright repeating FRBs are at distances that are amenable to the proposed experiment, \tmp{with the number of confirmed repeating FRBs constituting 2.6\% of all FRB sources discovered by the CHIME collaboration \citep{2023arXiv230108762T}}.  Most of these are inferred to repeat on a timescale of a few hours \citep{2019ApJ...885L..24C, 2023arXiv230108762T}, and they are known to show spates of intense activity -- FRB20201124A was clocked at several registered bursts per hour \citep{10.1093/mnras/stab3218}!  As shown below, the repeating bursts generally have flux densities at $\sim 1\,$GHz that our fiducial constellation parameters of four satellites each equipped with 10~m dishes would be sensitive to if they have similar flux densities at the targeted frequency. 
Furthermore, if the FRB is detected on Earth by a more sensitive instrument, this voltage time series could be used to effectively enhance the SNR as the geometric mean of the two instruments.

While we are hopeful that a sufficient sample of repeating FRBs will be found for the proposed experiment, it could also detect FRBs that occur in its field of view. We estimate here the FRB detection rate assuming that the array is not assisted by a more sensitive terrestrial telescope. The minimum specific fluence the proposed constellation is sensitive to is
\citep[e.g.][]{burke09}:
\begin{eqnarray}
    F_{\nu}^{\rm min} &=& \frac{8 k T_{\textrm{sys}} \tau}{{\pi D^2} N_{\rm sat} \sqrt{\Delta \nu \tau}} \,\text{SNR}_{\rm tot},\\
    &=& 1.4\,\textrm{Jy~ms}
    \left(\frac{D}{10~\textrm{m}}\right)^{-2} \left(\frac{N_{\rm sat}}{4}\right)^{-1}
    \left(\frac{\Delta \nu}{1~\textrm{GHz}}\right)^{-1/2}
    \left(\frac{\tau}{1~\textrm{ms}}\right)^{1/2} \left(\frac{T_\text{sys}}{20~\text{K}} \right) \left(\frac{\text{SNR}_{\rm tot}}{8} \right),
\end{eqnarray}
where $D$ is the diameter of each dish, $N_{\rm sat}$ is the number of satellites,
$\Delta \nu$ is the bandwidth, $\tau$ is the effective duration of the FRB, $F_\nu$ is the specific fluence, $T_{\textrm{sys}}$ is the system temperature of the receiver, and $\text{SNR}_{\rm tot}$ the total signal to noise at which the array detects the FRB.  At gigahertz frequencies,
 $T_{\textrm{sys}}\approx 20$ K is commonly achieved \citep{burke09}. Such a radio
telescope would have a beam with a solid angle of
$    \Omega_{\rm beam} = {4 c^2}/({\pi D^2 \nu^2}) = 5\times10^{-5}~\textrm{sr} \left({D}/{10~\textrm{m}}\right)^{-2} \left({\nu}/{5~\textrm{GHz}}\right)^{-2}$,
although, \tmp{at the expense of large data rates and computation that may not be possible in space}, modern radio dishes can be phased to generate hundreds beams, $N_{\rm beam}$, for a total solid angle of $\Omega_{\rm tot} = N_{\rm beam} \Omega_{\rm beam}$.

The all-sky rate of FRBs can be modeled as a power law where the rate above some
fluence limit is given by $R(> F_{\nu}) \propto F_{\nu}^{-\gamma}$. The CHIME survey has inferred $800$ FRBs occur per day above 5 $\rm Jy\cdot ms$ with $\gamma = 1.4\pm 0.2$ \citep{CHIMEFRB:2021srp}, near the Euclidean-space scaling of $\gamma =3/2$. Adopting the Euclidean scaling, the rate of detections for our antennas is 
\begin{align}
    R &= 800~\textrm{day}^{-1}~K_\nu^{3/2}~\left(\frac{\Omega_{\rm tot}}{4\pi}\right)~\left(\frac{F_{\nu}^{\rm min}}{5~\textrm{Jy}\cdot \text{ms}} \right)^{-3/2}, \\
      &= 2.0~\textrm{day}^{-1}~~ K_\nu^{3/2}
    \left(\frac{N_{\rm beam}}{100} \right)\left(\frac{D}{10~\textrm{m}}\right) \left(\frac{N_{\rm sat}}{4}\right)^{3/2}
    \left(\frac{\nu}{5~\textrm{GHz}}\right)^{-2}
    \left(\frac{\Delta \nu}{1~\textrm{GHz}} \frac{1~\textrm{ms}}{\tau}\right)^{3/4}  \left(\frac{T_\text{sys}}{20~\text{K}}\frac{\text{SNR}_{\rm tot}}{8} \right)^{-3/2},
    \label{eqn:R}
\end{align}
where $K_\nu$ is a bolometric correction factor that corrects on-average the specific fluence for the observation as $\nu$ relative to the CHIME $400-800~$MHz band.  Observations of the first repeater FRB~20121102 at 8\,GHz show similar flux densities to those at $1\,$GHz \citep{gajjar18}. \citet{2022arXiv220713669B} found a scaling of $F_\nu\propto \nu^{-1}$ for the repeater FRB~20180916B over $4-8~$GHz (but with fewer repetitions at these high frequencies relative to at $\sim 1\,$GHz) and a similar scaling is seen up to $6\,$GHz for the repeater FRB~20190520B \citep{2022arXiv220211112A}. The scaling $F_\nu \propto \nu^{-1.5\pm 0.3}$ has been measured between $1.1$ and $1.4~$GHz for the average spectrum of a large number bursts  \citep{macquart2019}.  Even with the strong $-1.5$ scaling such that $K_\nu= 0.1$ for a $\nu = 5\,$GHz observation, a rate of $R \approx 20\;$yr$^{-1}$ is anticipated for the fiducial specifications for which the parentheses in Equation~\ref{eqn:R} evaluate to unity.  This suggests that a non-negligible rate is not infeasible for $D>10$m if $N_{\rm beam}$ that have been achieved on the ground are possible, despite the relatively high frequencies of $\nu \gtrsim 3\,$GHz that we find are required owing to Milky Way scattering. 

The design also does not have to be radio dishes. Each satellite could instead be a plane of dipoles or small dishes. If the effective collecting area is comparable to $\pi D^2/4$, this would enable the same sensitivity as our dish but allow up to $\Omega_{\rm tot} \sim 4\pi$ and a huge number of FRBs per day! The additional cost to this approach is the computation required to phase all the dipoles to point simultaneously in hundreds or thousands of directions. \tmp{This would require onboard computation that is well beyond that on modern spacecraft.}

 If an individual satellite triggers on a burst, then it only needs to save the electric field time series over the $\sim 1\;$ms duration of the FRB (or instead the $\sim 0.1\, (5 \,\text{GHz}/\nu)^{2}$~s duration if the dispersive sweep is not corrected).  However, if the burst is not flagged by the satellite, the satellite either needs to record the electric field time series for longer than the time it takes to communicate with Earth-based antennas, which would signal that an FRB had occurred.  This approach requires the satellite to store at least $N_\text{beams} r/2\times   2 \Delta \nu = 10^{13} \,N_\text{beams} \,(r/10\,\text{AU}) (\Delta \nu/1 \,\text{GHz})$ numbers, each stored with at least one bit. (Even one bit is common in VLBI, but comes with some signal loss; \citealt{2017isra.book.....T}.)  While this storage demand is likely prohibitively high for $N_\text{beams} \gg 1$, if the signal is flagged for a SNR threshold on a single antennas, then the storage cost can be greatly reduced by a factor of $\sim \exp(-\text{SNR}^2/2)$ until all storage is actual FRB signal.   The $5~$GHz electric field of a millisecond-long FRB requires relaying back just $\sim 10^7$ numbers for correlation.
%
%



\section{Positional and temporal calibration}
\label{sec:calibration}

\subsection{Sources of acceleration and calibration timescales}

The outer Solar System is a lower acceleration environment than the medium-Earth orbit of most global positioning satellites. Even at medium-Earth orbit, global navigation satellites positions (ephemerides) are predicted to meter accuracy one day in advance, allowing updated ephemeris information to be uploaded to GNSS satellites only every several hours \citep{misraenge}.  A primary factor limiting the precision of GNSS satellite ephemerides is various stochastic accelerations, including variations in the Sun's irradiance. Stochastic acceleration sources also set the timescale that the satellite ephemerides in the proposed experiment need to be calibrated, which in turn inform the types of calibration systems that would be required. In what follows, we estimate the magnitude of different sources of stochastic accelerations for satellites in the outer Solar System.  Sources of gravitational acceleration are the most concerning, as an active system that adjusts in response to an accelerometer the satellite orbit can potentially correct for non-gravitational accelerations.

\begin{description}

\item[gravity from asteroids] 
Accelerations from the passages of asteroids that are close enough so they occur over a much shorter timescale than their orbital time are the main concern for calibration.  Let us consider the requirements to have a sufficient number of close passages for calibration to be necessarily on a timescale of $t_{\text{calib}}$. The shorter $t_{\text{calib}}$, the more challenging for calibration.  In order to require calibration every $t_{\text{calib}}$, sufficiently massive asteroids within a distance $v \, t_{\text{calib}}$ must be commonplace, where $v$ is the satellite--asteroid relative velocity.  For that asteroid to displace the satellite by $1~$cm over a time $t_{\text{calib}}$, it must have a mass satisfying $m > m_{\rm cr} \equiv v^2 \times 1 \, \text{\,cm}/G$, where $G$ is Newton's constant or, assuming a mass density of 2~g~cm$^{-3}$ and a spherical body, a radius of at least $3 \, (v/1 \, \text{km~s}^{-1})^{2/3}~$km.  Here $v\sim 1 \, \text{km~s}^{-1}$ is motivated as, if our satellites are in orbit, we expect passage at a fraction of the $\sim 5~$km~s$^{-1}$ Keplerian velocity at tens of AU radii owing to the moderate eccentricities of asteroids and possibly of our satellites.  However, our favored configuration would be unbound satellites that are drifting out at $10-100\,$km~s$^{-1}$, in which case gravitational accelerations from asteroids would be even much smaller and due to larger masses.  For asteroid deflections to be commonplace requires a number density of at least $n_{\rm KB} = (v t_{\text{calib}})^{-3}$ or a total number of $N_{\rm KB} \approx 4\pi/3 \,r_{\rm KB}^3 \sin(\theta_{\rm KB}) n_{\rm KB}=  4\times 10^{12}  (r_{\rm KB}/40 \, \text{AU})^3(v/1 \, \text{km~s}^{-1})^{-3} (t_{\text{calib}}/ 1 \,\text{week})^{-3} \sin(\theta_{\rm KB})$ asteroids above $m_{\rm cr}$, where $\theta_{\rm KB}$ describes the asteroids' angular extent above and below the ecliptic.  For $\theta_{\rm KB} = 10^\circ$ and $r_{\rm KB}= 40\,$AU characteristic of the Kuiper belt and an orbital relative velocity of $v=1\,$km~s$^{-1}$, an anomalous acceleration that results in a centimeter displacement over a week requires a number density of kilometer-sized asteroids that is 4-5 orders of magnitude higher than in models that match observations \citep{2004AJ....128.1916K,Schlichting2009}. (Optical observations probe asteroids with radii of $\gtrsim 10~$km, with X-ray diffraction constraining the abundance at $\approx 200$\,m.)   Similarly, the total mass we require is $N_{\rm KB} m_{\rm cr}  >100\, M_\earth \; (r_{\rm KB}/40~\text{AU})^3 (v/1 \, \text{km~s}^{-1})^{-1} (t_{\text{calib}}/ 1 \, \text{week})^{-3}~\sin(\theta_{\rm KB})$, much larger than the $\sim 0.01-0.1~M_\odot$ estimated mass for the Kuiper belt \citep{2004AJ....128.1364B}.  If we instead calibrate to the $N_{\rm KB}\sim 10^{8-9}$ asteroids with radii of $>3\,$km found in models that are matched to Kuiper belt constraints \citep{2004AJ....128.1916K,Schlichting2009}, then following the above logic, this indicates a calibration time of $t_{\text{calib}} \approx 10~$weeks.  Thus, we estimate that the calibration time for a satellite on a bound orbit \emph{within} the Kuiper belt is $\sim 10~$weeks, and this time should be longer if the satellite resides elsewhere or is unbound. 

\item[gaseous drag]  The drag force on a surface of effective area $A_{\rm eff}$ traveling through the Solar wind with densities of $n = 0.05 \,(r/10 \,\text{AU})^{-2}\;$cm$^{-3}$ and particle velocities of approximately $v= 500$\,km\,s$^{-1}$ \citep{1994PhDT.........1V} can be computed as $F_{\rm drag} = 1.4 \, m_p \, n \, v^2 \, A_{\rm eff}$. Drag results in a 1\,cm displacement after a time of $t_{\rm drag} = \sqrt{2 M_{\rm sat} \times 1 \, \text{cm}/F_{\rm drag}} =  9 ~(A_{\rm eff}/10 \text{m}^2)^{-1/2} (M_{\rm sat}/10^3\, {\rm kg})^{1/2} (r/30\,\text{AU})$~days. The Voyager spacecrafts show factor of $\sim 2$ variations in the density of the Solar wind on the 27~day Solar rotation period  \citep{1994PhDT.........1V, 2003GeoRL..30.2207R}, which would result in a somewhat longer time for the more relevant stochastic drag relative to our homogeneous estimate.  Neutrals from interplanetary space penetrate the Solar System with the local interstellar abundance of $0.1\,$cm$^{-3}$ at a velocity of $30\,$km~s$^{-1}$ \citep{1977RvGSP..15..467H}.  Their drag results in a 1\,cm displacement over $t_{\rm drag} \approx 40$\,days, becoming the dominant drag at $r \gtrsim 100\,$AU. 



\item[Lorentz force on a charged spacecraft] The spacecraft will inevitably build up a charge, especially since the radio dish is a large conducting element. The maximum possible charge is roughly set by that which is sufficient to repel Solar wind protons from striking and further charging the spacecraft, namely $m_p v^2 = Z_{\rm max} \, e^2/R$, where $v$ is the velocity of intersecting particles -- the $500\,$km~s$^{-1}$ velocity of the Solar wind -- and $R$ is the effective extent of the satellite.  The Lorentz force, $Z_{\rm max} \, e \, (v_{\rm sat}/c) \, B $, acting on the satellite traveling through an interplanetary magnetic field with $B = 1\,\mu\,G$ \citep[and with factor of $\sim 2$ monthly variations]{1994PhDT.........1V} results in a displacement of $1\,$cm after $14\, (R/10 \,\text{m})^{-1/2} (M_{\rm sat}/10^3\, {\rm kg})^{1/2}~(v_{\rm sat}/50 \,\text{km~s}^{-1})^{-1/2}  (B/1\, \mu G)^{-1/2}\,$days.  Spacecraft charging would be less important outside of the heliosphere owing to the much smaller velocities of the incident plasma.

\item[solar radiation pressure]  Solar radiation pressure variations are a primary reason for the meter ephemeris errors on day timescales in global positioning satellites \citep{10.1007/BFb0011321}.  Even though the radiation pressure is down by a factor $10^{-3}(r/30 \, \text{AU})^{-2}$ in the outer Solar System relative to our terrestrial environment, the centimeter ephemeris accuracy requirement is two orders of magnitude more stringent. During active periods, the Sun shows 0.2\%  peak-to-peak variations on the timescale of its 27 day rotation period  \citep{2004A&ARv..12..273F}.  If we parameterize the correlation function of the variations as $|\xi(t) - \xi(0)|= 10^{-3} \, (t/13.5 \,\text{days})^{1/2}$ --motivated by the scaling of the power spectrum of these fluctuations on $1-10~$day timescales \citep{2004A&ARv..12..273F} --, we find that fluctuations in the solar irradiance will result in an acceleration of $ \sim  |\xi(t) - \xi(0)|\, F_\odot A_{\rm eff}/M_{\rm sat}$ and, integrating this acceleration with respect to time, a $1$~cm displacement of the satellite every $\sim 11~(A_{\rm eff}/ 10~\text{m}^2)^{-0.4} (M_{\rm sat}/10^3 ~\text{kg})^{0.4} (r/30\, \text{AU})^{0.8}~\text{days}$, where $F_\odot$ is the momentum flux of the Sun and $A_{\rm eff}$ is the effective surface area of the satellite.  Understanding thermal re-emissions of the absorbed light is also important.
\item[dust collisions] Collisions with dust particles will accelerate the spacecraft.  The Ulysses and Voyager spacecrafts found that outside of $5\,$AU the population of $\sim 0.1\,\mu$m dust was dominated by an interstellar population with impact speeds of $v_{0.1\mu} = 30\,$km~s$^{-1}$ and a resulting flux of $\approx 10^{-8}$cm$^{-2}\,$s$^{-1}$ \citep{1994A&A...286..915G, 2005LPICo1280...63G}.  Such fluxes correspond to $N_{\rm coll}\approx 100 \,(A_{\rm eff}/10 \,\text{m}^2)$ dust particles striking an effective area of $A_{\rm eff}$ per day, assuming the spacecraft velocity is $\lesssim v_{0.1\mu}$.  The stochastic (e.g. $1/\sqrt{N_{\rm coll}}$) component of the acceleration is so small that it takes hundreds of years to result in a 1 cm displacement for the Ulysses-inferred average dust mass of $3\times10^{-14}\,$g and a satellite mass of $M_{\rm sat} =10^3\,$kg.  More important are the larger dust grains seen in zodiacal light.  While outside of several AU the reflectance of these particles contributes negligibly to this diffuse emission and hence the distribution of such tens-of-micron-sized particles is unconstrained, within this radius it is found that an order unity fraction of zodiacal particles is relatively isotropic with density scaling radially as $\propto r^{-1.3}$ \citep{grun}. For the following, we assume that this radial scaling still holds well beyond several AU. If a $30\mu$m grain with mass $8\times10^{-7}$g is conservatively assumed with an abundance normalized to match the zodiacal background \citep[$2\times10^{-17}$cm$^{-3}$ at $5\,$AU;][]{1994A&A...286..915G}, one particle would strike the satellite every $1.4 \, (A_{\rm eff}/10\, \text{m}^2)^{-1} (r/10 \,\text{AU})^{1.3}  (v_{\rm sat}/100 \,\text{km s}^{-1})^{-1}$~days.  A single particle would result in an acceleration that leads to a 1~cm displacement after $1.4 \, (M_{\rm sat}/10^3\, {\rm kg}) (v_{\rm sat}/100 \,\text{km s}^{-1})^{-1}$~days.  These estimates for the timescale for zodiacal dust are somewhat smaller than the other acceleration timescales considered above, but we have evaluated with very conservative values (a high $v_{\rm sat}$, a small $r$ and a large mass per particle\footnote{The number density of dust grains per log mass at an $1\,$AU is inferred to scale with particle mass as $m^{-0.5}$ \citep{grun}.  Our estimate assuming all particles are in very large $30\,\mu$m grains overestimates the acceleration.}). Thus, we suspect that dust collisions again require calibration on a week or longer timescales.

\item[time dilation from Solar System masses]
The mass distribution in the Solar System also affects onboard clocks. Clocks run a factor of $ (1- 3 r_M/[2r])$ slower when in orbit at distance $r$ around an enclosed (spherically symmetric) mass of $M$ relative to the global Schwarzschild time, where $r_M \equiv GM/c^2$ is the gravitational radius, and the time dilation is similar for unbound trajectories.  
For a bound orbit, a tenth of an Earth mass uncertainty in the enclosed mass, near the upper limit on the mass in the Kuiper belt, would would result in a clock drift of $\Delta t_{\text{clock}} = 0.03~(t/1 \,\text{week}) (100 \,\text{AU}/r)\,$ns after a time $t$.   Any uncertainty in the enclosed mass would be easy to calibrate as the timescale for this drift is long. Consider instead a passing outer Solar System object of mass $M$ and distance $d_M$ -- such as some Kuiper belt object. Its mass would contribute a clock drift rate of $ \sim r_M/d_{M}$. 
Using that the cumulative mass distribution of Kuiper belt objects above mass $M$ is found to be $N(M) \sim M^{-2}$ \citep[e.g.][]{Schlichting2009}, we expect an object within a logarithmic mass interval in $M$ to come within a distance of $d_M = N(M)^{-1/3} = M^{2/3}$.  Thus, $M/d_M = M^{1/3}$, suggesting that the largest bodies create the largest timing errors.  A worst case scenario might be an Earth-massed object at a distance $\ell_{\rm sep} = 1\,$AU from one detector, which would result in a detectably large drift of $\sim 3$~ns~day$^{-1}$.  Of course, this long-timescale drift (or even ones that are hundreds of times smaller) should be detected in calibration, and, hence, the $M/\ell_{\rm sep}$ of the object measured!  Clock drifts on week or month timescales from the gravity of (small) asteroids that make close enough approaches are negligible.

\end{description}
In conclusion, variations in Solar irradiance, dust collisions, gaseous drag, and possibly charging require calibration on $\gtrsim 10~$day timescales for motivated specifications for the proposed satellite constellation. Gravitational accelerations that result in a 1\,cm displacement occur over longer timescales.  Gravitational clock drifts owing to passing bodies that occur over weeks or months should be negligible.

\subsection{Calibration methods}


Measuring time delays to within (5\,GHz)$^{-1}$ = 0.2\,ns implies that the relative detector positions must be modeled
to six centimeters. We suggest two possible strategies for achieving such calibration. First, our favored calibration strategy is direct trilateration between the satellites themselves. Our second calibration method would combine satellite--Earth communications with calibration off of astrophysical sources.  

We first consider a calibration strategy that relies on GNSS-like trilaterations.  On Earth, these techniques regularly provide meter accuracy, if not centimeter to millimeter with longer integrations that use the carrier phase: Generally four satellites are needed for geolocation, but, if the user possesses an accurate clock, only three are required.  
In this calibration mode, the proposed experiment would be a much enlarged version of GNSS, where members of the constellation would relay signals to measure each other's distances.  Modulo signal to noise effects, enlargement itself does not affect the distance accuracy. 

First, let us consider the power requirements. The proposed experiment equips each satellite with a radio dish, which is fortunate as this is also necessary for the dishes themselves to broadcast strong enough signals for ranging. Ranging works by modulating the carrier wave at frequency $\nu_{\rm car}$ with some code with frequencies of $\sim \delta \nu$, which when multiplied by the carrier approximately sets the bandwidth.  For global positioning satellites, $\delta \nu\sim 1\,$MHz is chosen to reduce computation and radio frequency interference.  The receiver then uses a matched filter with different temporal delays to find the arrival time.  The error in the ranging delay from this `delay-lock-loop' procedure is (e.g.~\citealt[][see their section 10.5]{misraenge})
\begin{eqnarray}
\sigma_t^{\text{ranging}} &=& T_c \sqrt{\frac{ \epsilon \, k_b \, T_{\rm sys}}{4 \, t \, P_{\rm rec}}}, \label{eqn:dtranging} \\
&=& {0.04\;} \text{ns\;}   \left(\frac{10\,\textrm{MHz}}{T_c^{-1}}\right)  \left(\frac{10\,\textrm{m}}{D}\right)^{2} \left(\frac{5\,\text{GHz}}{\nu}\right)
   \left(\frac{x}{100 \, \rm AU} \right) 
    \sqrt{ \left(\frac{\epsilon}{0.1} \right) \left(\frac{1\,\textrm{min}}{t}\right)   \left(\frac{100\,\text{Watt}}{P_\text{ant}} \right) \left(\frac{T_\text{sys}}{20\,\text{K}} \right)},\nonumber
\end{eqnarray}
where we have used that the received power can be written as  $P_{\rm rec} = \pi D^2  P_{\rm ant}/(4 \Omega_{\rm beam} x^2)$, where $P_{\rm ant}$ is the power broadcast by the other antennas, and that the noise power is white with $P_N = B k_B T_{\rm sys}$, where $B$ is the bandwidth. Here $t$ is the integration time, and $T_c$ is the duration of code chips such that $T_c \sim 1/(2\delta \nu)$. (The code for GNSS applications are `chips' of zeros and ones with duration $T_c$.  We are borrowing this approach.)  The parameter $\epsilon$ is the correlator spacing for the delay lock loop -- the circuit that attempts to phase up the satellites matched code with the received signal.  GNSS applications have to filter for radio frequency interference and this generally restricts $\epsilon >0.1$ \citep{misraenge}, but it is likely in our case that smaller $\epsilon$ may be achievable with the absolute minimum being $\epsilon \approx (B T_c)^{-1}$ \citep{braasch}.\footnote{One can approximately derive Equation~\ref{eqn:dtranging} from optimizing the location of a matched filter.  In ranging applications, the correlation function of matched code with the signal offset by a time $ \tau$ is chosen so that it is the triangle function scaling as $(1 -  \tau/T_c)^2$, as happens for a random sequence of chips.  This results in a likelihood function  ${\cal L}$ of  $-2 \log {\cal L} = P_{\rm rec}(1 -  \tau/T_c)^2/ (k_b T_{\rm sys } B)  \times N$, where $N = 2 Bt$ is the number of temporal samples and we have assumed one polarization.  The standard deviation which $\tau$ can be measured is then $\sigma_t^{\text{ranging}} = (-\partial_{\tau}^2 \log {\cal L})^{-1/2}$, which yields the result to the factor of $\sqrt{\epsilon}$ or so. 
We suspect one reason for the $\sqrt{\epsilon}$ improvement over the `ideal' matched filter is that the delay-lock-loop circuit has access to all lags and not the just a grid of temporal samples that this matched filter estimate assumes.} The above is for ranging using the low-frequency code.  Once the satellite is localized to better than $c/\nu_{\rm car}$ so that integer wavelength ambiguities are resolved, the carrier wave can then be used to measure the distance.  The carrier provides a precision that is even much higher and likely limited by other considerations than SNR \citep{carrierpahse}.



However, it may also be possible to use a network of sources to calibrate the distances rather than direct trilateration using our satellite network.  
Since FRBs are likely the only extragalactic radio object that appear point-like at Solar System-scale baselines (\S~\ref{sec:sources}), they are also likely the only extragalactic source that is useful for calibration.  FRBs known to be more distant than the distance reach of the array can be used for calibration (but also FRBs with well-characterized distances). Each distant FRB provides $N_D-3$ constraints after fitting for the source's angle on the sky and the FRB event time, where $N_D$ is the number of detectors.  Repeating FRBs are even more valuable, with each repetition providing $N_D-1$ additional constraints once the relative motion of the repeating FRB is determined.  The goal of calibration is to constrain the $4 N_D$ satellite space-time coordinates.  In the case $N_D =4$ (6), to calibrate the network requires $16$ ($8$) one-off FRBs over a calibration time or, instead, 6 (5) repeating FRBs.  The calculations in \S~\ref{sec:FRBrate} suggest that it may be challenging to detects 5-16 FRBs within the $\sim 1~$week calibration time needed if non-gravitational sources of acceleration cannot be removed.  If so, a system that relies predominantly on distant FRBs for calibration would need a precise accelerometer to correct for non-gravitational accelerations.\footnote{\tmp{FRBs are known to be highly polarized \citep{petroff21}, which potentially provides an additional lever for calibration.  The polarization should be the same for all the detectors as the differential Faraday rotation on Solar System-scale baselines is likely negligible; Faraday rotation is dominated by large-scale coherent magnetic fields.}}

Similarly, pulsars may be the only common point-like Galactic source.  See \S~\ref{sec:otherscience}, as Galactic pulsars may not be sufficiently point-like, especially for the longer baselines we consider.  However, calibration off of pulsars would be limited to the small fraction of pulsars at high Galactic latitudes to avoid ISM scattering.



The clocks on the satellites in our constellation must be kept synchronized.  The clock requirements for centimeter space-time precision and weekly calibration are somewhat more stringent than those for the atomic clocks on GPS satellites, which err at several nanoseconds a day rather than the few tenths of a nanosecond per week that would be required to not be the dominate error in the ephemerides.  Such clock specifications are somewhat more stringent than the $4~$ns per $23$ days accuracy achieved by the Deep Space Atomic Clock mission, a miniature atomic clock that was launched in 2019 to demonstrate improved space-clock technology for future NASA missions as well as for next-generation GPS \citep{2021Natur.595...43B}. 
The best atomic clocks in the world exceed the performance of the Deep Space Atomic Clock by many orders of magnitude \citep{RevModPhys.87.637}.  Less accurate clocks than the tenths of nanosecond per week specification entails more regular calibration or $\gtrsim1\,$cm ephemerides errors.

\section{Distance measurement accuracy}
\label{sec:discussion}
\begin{figure}
\begin{center}
\includegraphics[scale=0.85]{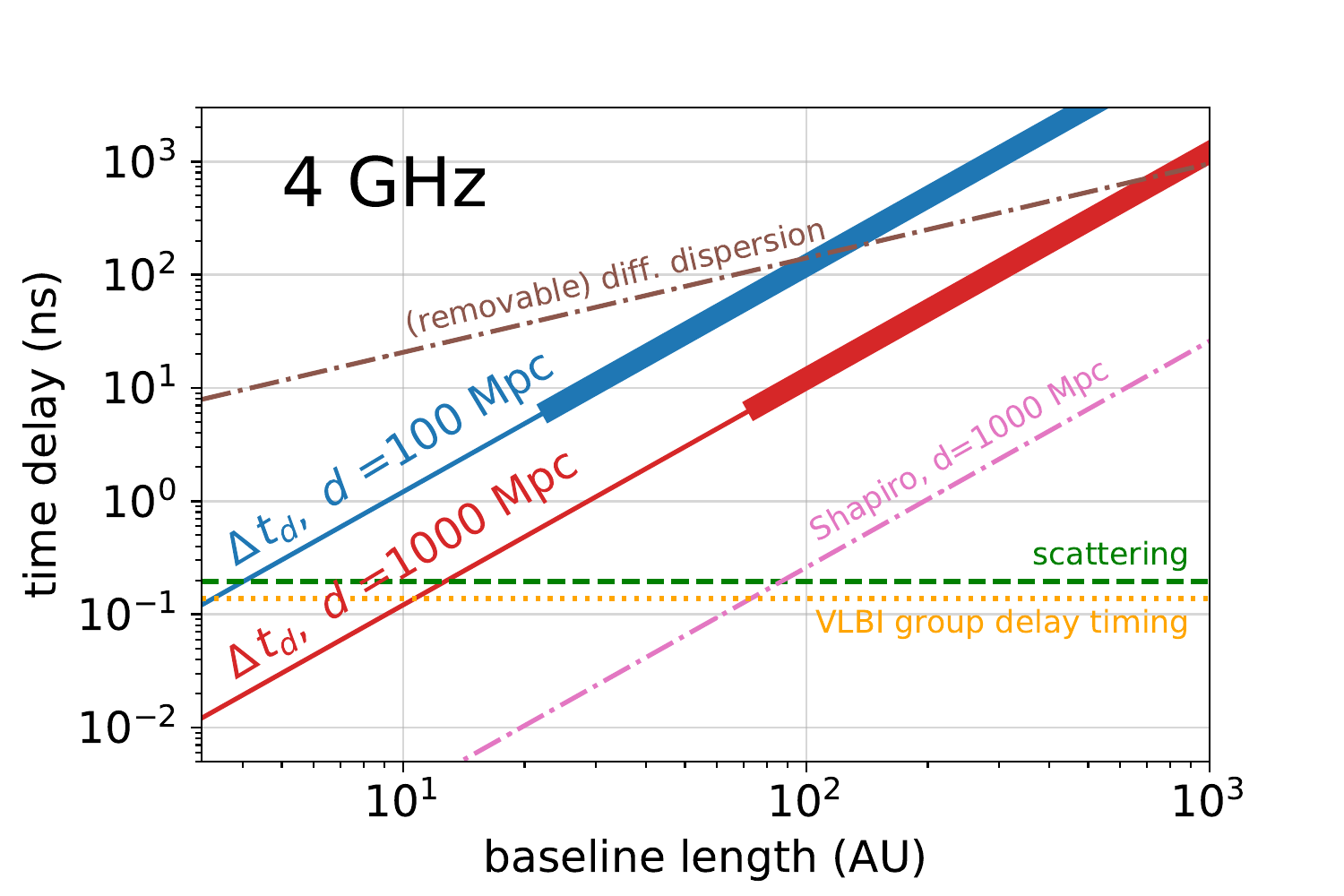}
\includegraphics[scale=0.85]{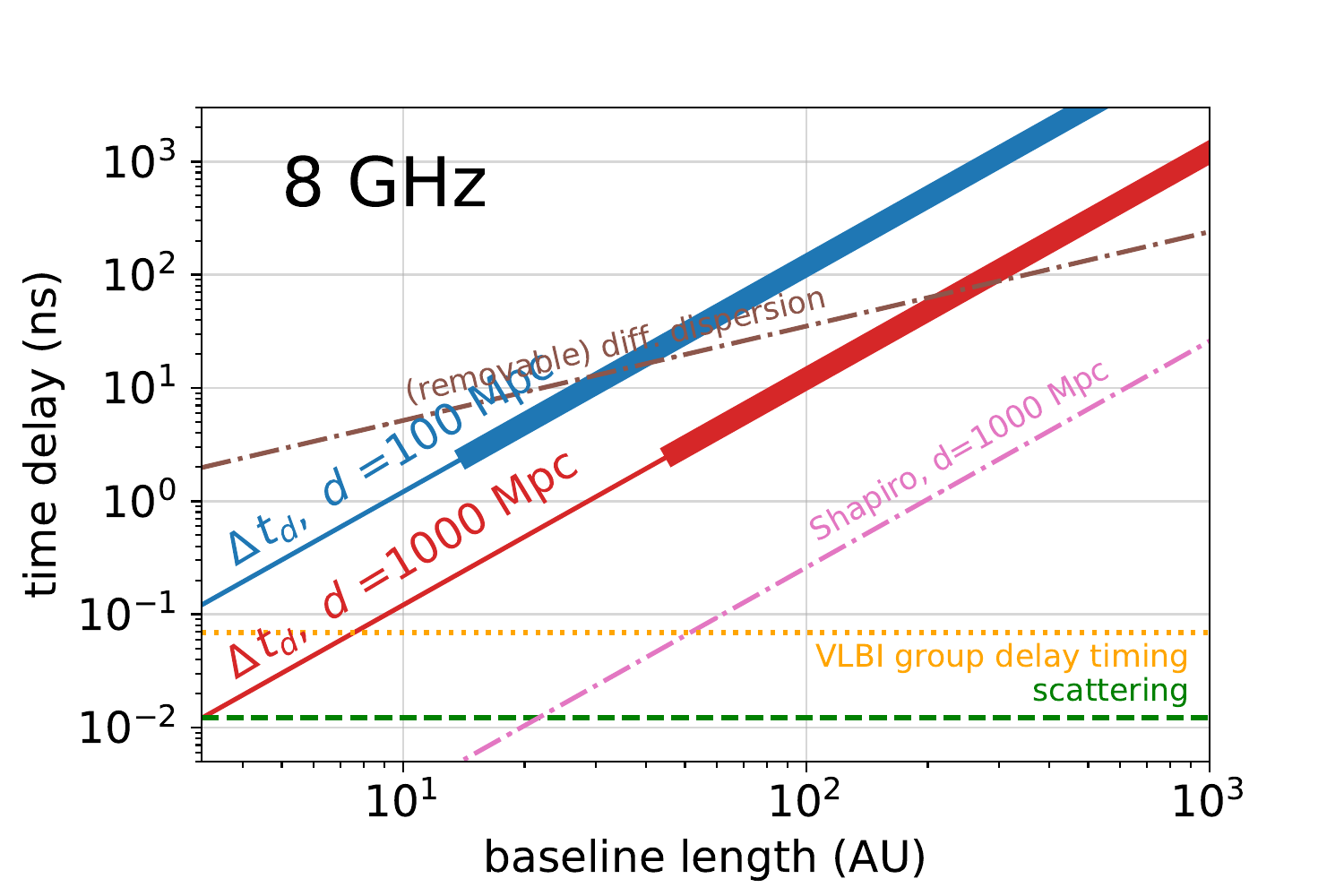}
\end{center}
\caption{Time delays versus projected baseline length $x_\perp$ for $\nu=4\,$GHz (top panel) and $\nu=8\,$GHz (bottom panel).  The solid lines are the cosmological delay owing to wave front curvature, $\Delta t_d$, for $d=100\,$Mpc (blue) and $d=1000\,$Mpc (red).  The other lines are estimates for different systematics: the differential dispersion from the Galaxy (Equation~\ref{eqn:sigmadisp} assuming SM$\equiv C_n^2 L = 10^{-4}\; \textrm{kpc~m}^{-20/3}$), the scattering time (taken to be $\sigma_t^{\rm sc} = 50\,\text{ns}\,[\nu/1\,\text{GHz}]^{-4}$, $2\times$ larger than the mean NE2001 model at $b = 90^{\circ}$; Equation~\ref{eqn:sc}), the Shapiro time delay for $d=10^3$Mpc (assuming the cosmological delay is dominant; Equation~\ref{eqn:tgravquadcosmo2}), and the VLBI group delay timing error (assuming SNR$=5$ and $\Delta \nu=0.2\, \nu$; Equation~\ref{eqn:sigmat}).  Even though it is often larger than the signal $\Delta t_d$, differential dispersion can be removed using its frequency dependence (\S~\ref{sec:dispersion}).  The baselines over which the solid lines representing $\Delta t_d$ are thickest show where VLBI timing and scattering delays (all calculated in the manner described above) are each $<3\%$ of $\Delta t_d$.  Shapiro delays, if not removed using their quadrupolar shape, limit $<3\%$ precision to $d \lesssim 10^3~$Mpc.   
}  
\label{fig:withbaslein}
\end{figure}

The previous discussion motivates that the distance-dependent time delays $\Delta t_d$ may be measurable with a Solar System-scale interferometer with an uncertainty that can approach $\sim \nu^{-1}$ or even better.
Figure~\ref{fig:withbaslein} shows the various time delays calculated in \S~\ref{sec:timing} versus projected baseline length $x_\perp$ for $\nu=4\,$GHz (top panel) and $\nu=8\,$GHz (bottom panel).  The solid lines are the cosmological signal of interest that owes to wave front curvature, $\Delta t_d$,  for $d=100\,$Mpc (blue) and $d=1000\,$Mpc (red).  The other lines are estimates for different systematic delays: the scattering time (taken to be $\sigma_t^{\rm sc} = 50\,\text{ns}\,[\nu/1\,\text{GHz}]^{-4}$,  $2\times$ larger than the mean NE2001 model at $b = 90^{\circ}$; Equation~\ref{eqn:sc}), the Shapiro time delay (assuming the cosmological delay is dominant; Equation~\ref{eqn:tgravquadcosmo2}), the VLBI group delay timing error  $\sigma_{t, \rm group}^{\rm VLBI}$ (assuming SNR$=5$ for the baseline with $d{\rm SNR}/d\nu$ uniform over a band of $\Delta \nu=0.2\, \nu$; Equation~\ref{eqn:sigmat}), and the differential dispersion from the Galaxy (Equation~\ref{eqn:sigmadisp} assuming SM$\equiv C_n^2 L = 10^{-4}\; \textrm{kpc~m}^{-20/3}$).  While the differential dispersion is even larger than our signal $\Delta t_d$ for the shorter baselines, it can also be removed owing to its frequency dependence at the cost of increasing the effective VLBI timing noise.  While the timing noise on the group delay is shown for the case of no differential dispersion, we find that the delay from the absolute phase when fitting out dispersion is half this value (see Appendix~\ref{ap:fittingdisp} for discussion).   Similarly the Shapiro time delay can be removed with a six satellite configuration, as this adds the two additional constraints needed to remove it using its quadrupolar shape.

Figure~\ref{fig:withbaslein} also shows how well cosmic distance could be constrained by the proposed experiment as a function of the typical baseline length. The baselines over which the solid lines representing $\Delta t_d$ have a thick line style indicate those for which the VLBI timing and scattering delays are each $<3\%$ of $\Delta t_d$.  Here we are using  $\sigma_{t, \rm group}^{\rm VLBI}$ and not half this value as we find is possible. For $\nu=4\,$GHz (top panel), the fiducial scattering delay is important and so limits $3\%$ precision to baselines of $>20$~AU for $d =100~$Mpc and $>70$~AU for $d =1000~$Mpc.  For some sightlines the scattering will be lower and hence a higher precision possible. At $8\,$GHz the scattering timescales are much smaller, and a $3$\% measurement of $\Delta t_d$ is possible for somewhat smaller baselines compared to $4\,$GHz. 

Both scattering and dispersion become less problematic with increasing frequency, whereas Shapiro time delays are frequency independent and limit the precision at longer baselines in a manner for which there is no benefit from increasing the baseline length unless they are removed.  For all baseline lengths, Shapiro delays limit $3\%$ precision to sources within $d \approx 1000~$Mpc.  The Shapiro time delay becomes the dominant noise for projected baselines of $x_\perp \gtrsim 100\,$AU.\footnote{Our simple calculation for the cosmological Shapiro delay does not include redshift evolution in the matter field, which likely leads to additional inaccuracy for $d \gtrsim 3000\,$Mpc.}

\begin{figure}
\begin{center}
\includegraphics[scale=0.55]{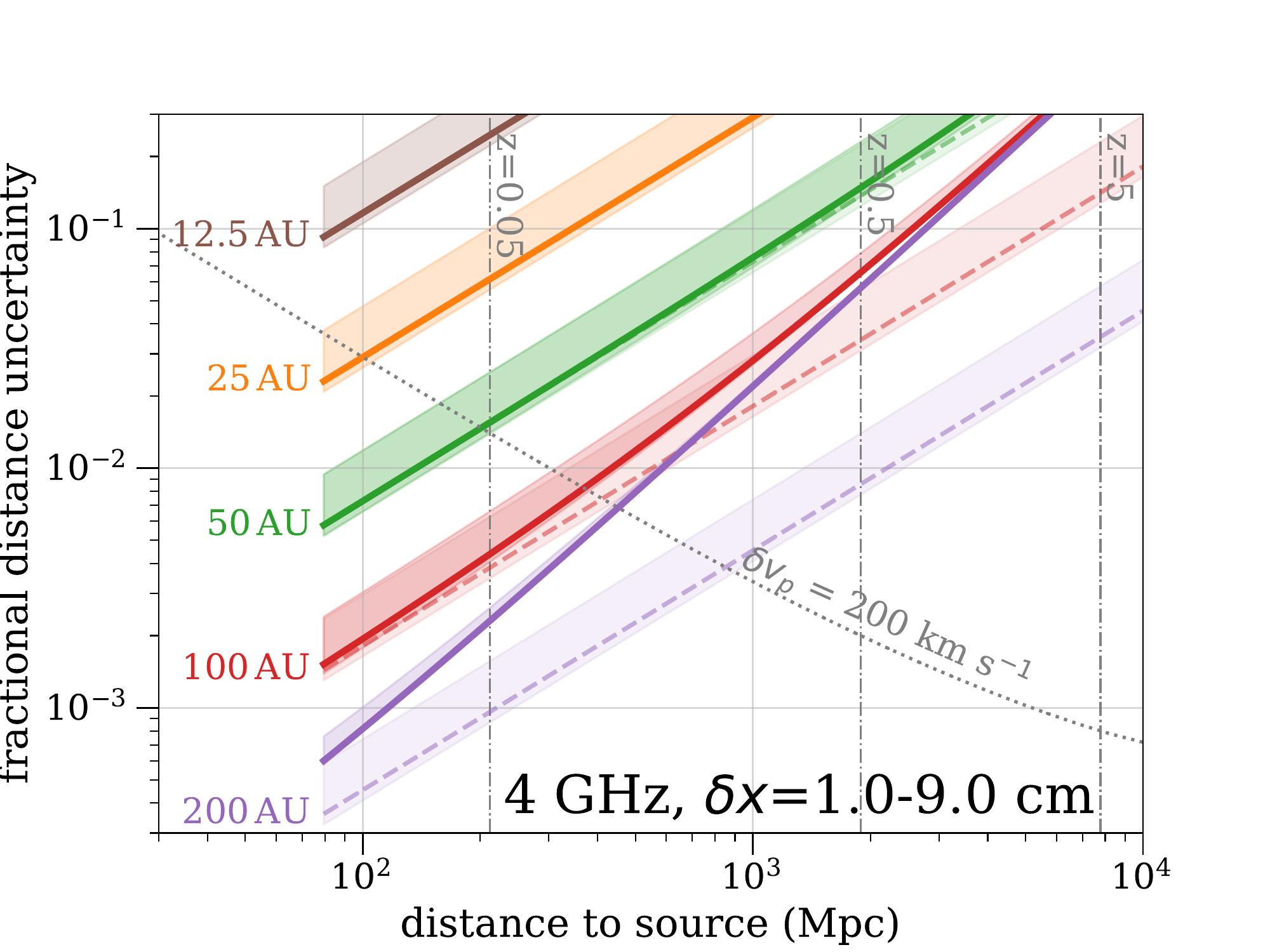}
\includegraphics[scale=0.55]{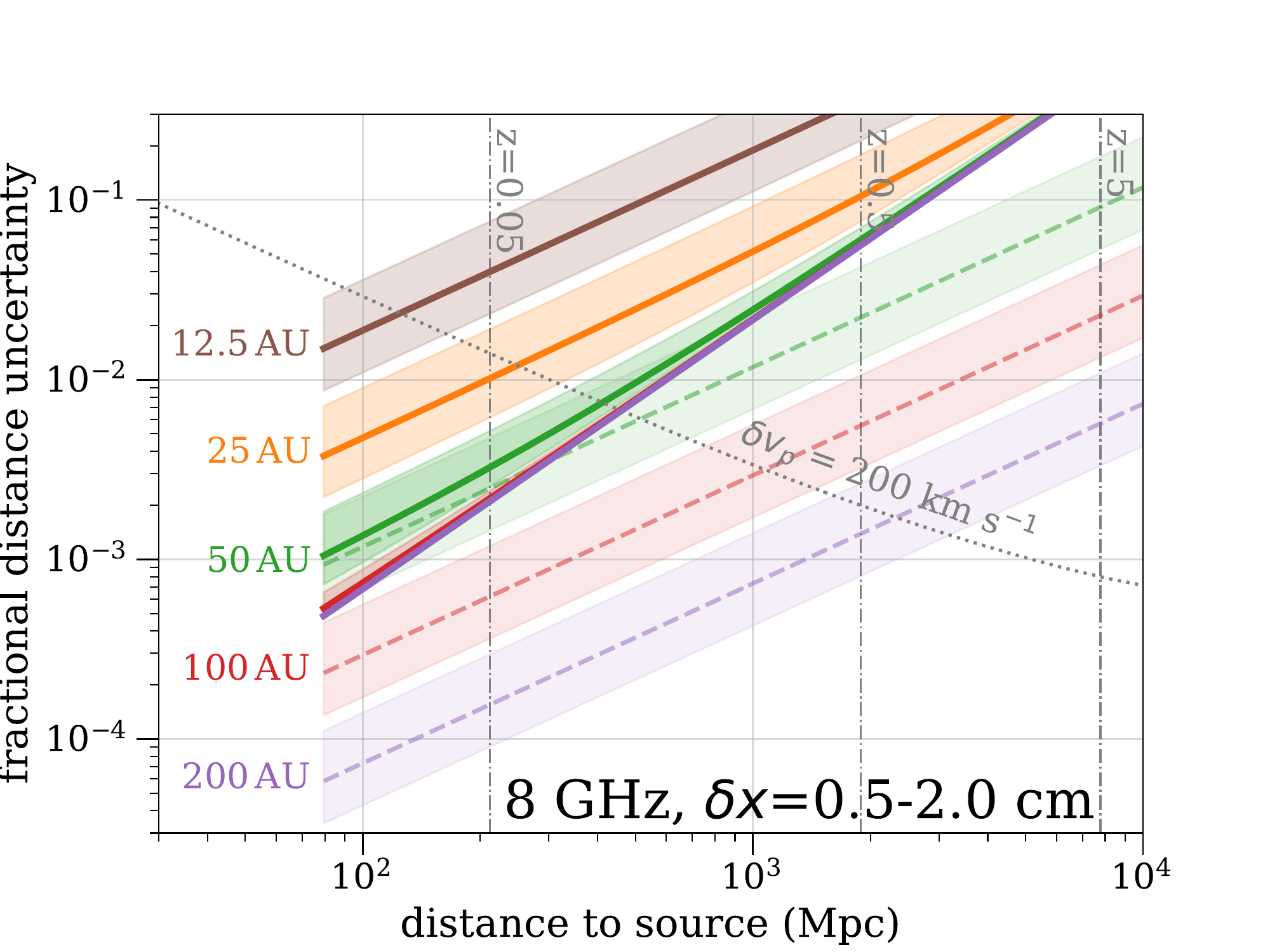}
\end{center}
\caption{\tmp{Estimated fractional distance error, $\sigma_d/d$, to a \emph{single} FRB as a function of distance $d$, for observations at $\nu=4\,$GHz (top panel) and $\nu=8\,$GHz (bottom panel).  The different colors indicate different detector separations $x_\perp$: 12.5~AU in brown, 25~AU in orange, 50~AU in green, 100~AU in red, and 200~AU in purple. The solid curves assume to compute the timing error ($\sigma_t$) that each baseline vector is constrained to a 1$\sigma$ error of $\delta x = 3\,$cm in the top and $\delta x = 1\,$cm in the bottom panel -- an error which includes the VLBI timing error, in addition to including the noise from scattering and Shapiro time delays assuming the same specifications as used in Figure~\ref{fig:withbaslein}.  The highlighted regions show how these uncertainties change over the specified range of $\delta x$.  The dashed curves, only shown for the three longest baselines, are the same but assume the Shapiro delay is removed owing to its different angular dependence.  The dotted curve is the fractional redshift uncertainty that arises if the peculiar velocity can be corrected for using large-scale structure observations to $200\,$km\,s$^{-1}$. The vertical lines show the comoving light-travel distance that corresponds to redshifts of $z=0.05,\,0.5$ and $5$ in a flat $\Lambda$CDM cosmology with $\Omega_M=0.3$ and $h=0.7$.}}
\label{fig:distance_uncertainty}
\end{figure}

These estimates for the timing noise relative to $\Delta t_d$ suggest that the proposed experiment would make cosmologically interesting distance measurements with a detector separation of at least $x_\perp \approx $~10\,AU. Figure~\ref{fig:distance_uncertainty} quantifies this further. It shows estimates for the fractional distance error, $\sigma_d/d$, to a single FRB as a function of their distance, for observations at $\nu=4\,$GHz (top panel) and $\nu=8\,$GHz (bottom panel). The fractional distance error is calculated using our estimate that $\sigma_d/d \approx \sigma_t/\Delta t_d$ when the $N_A-1$ equals the number of constraints, where $\sigma_t$ is determined by adding in quadrature the scattering, Shapiro, and position+VLBI timing error $\delta x/c$.  While the former two errors assume same specifications as described for Figure~\ref{fig:withbaslein}, we have combine the VLBI timing error and detector positional errors into the total geometric error $\delta x$.  For $\nu=4\,$GHz we have chosen $\delta x=3~$cm and for $\nu=8\,$GHz we have set $\delta x=1~$cm, and we note that the VLBI absolute phase timing error once marginalizing over dispersion is given by $c \,\sigma_{t, \rm group}^{\rm VLBI}/2 = 0.4\,\textrm{cm}\times(1~\textrm{GHz}/\Delta \nu_{\rm RMS}) (5/\textrm{SNR})$ and the effective timing error decreases when more than four detectors are used owing to the increase in baselines. 

The different curves indicate different detector separations $x_\perp$: 12.5~AU in brown, 25~AU in orange, 50~AU in green, 100~AU in red, and 200~AU in purple.  The highlighted regions vary $\delta x$ over a range of $1-10~$cm for $\nu=4\,$GHz and $0.5-2~$cm for $\nu=8\,$GHz.  Without Shapiro delay removal, four detectors with roughly these separations are required to reach the quoted precision (solid curves), and six are required for Shapiro delay removal (dashed curves).\footnote{\tmp{The VLBI timing error on the more optimistic six-detector configuration would be improved by a factor of $\approx 2.5$ if Shapiro delays are not fit owing to the increase in visibilities so that $\sigma_d/d= \sigma_t/\Delta t_d$ overestimates the error (\S~\ref{sec:geometric_dist}). Ignoring the Shapiro delays we find is appropriate at $x_\perp \lesssim 100~$AU and so the distance errors for this case could be even smaller than those quoted in this section}.}

The top panel in this figure considers $\nu =4\,$GHz.  The orange curves show our estimate for the error from a constellation with typical baselines of $x_\perp \approx 25\,$AU on a single FRB. This configuration would achieve  $\sigma_d/d \approx 0.05$ out to $d= 200\,$Mpc for our fiducial choice of $\delta x = 3~$cm.  A distance of $200\,$Mpc corresponds to approximately $z=0.05$ in the $\Lambda$CDM cosmology ($\Omega_M=0.3$ and $h=0.7$; Appendix~\ref{sec:curvedspace}), the median redshift of the SHOES supernova sample. For the $\nu =8\,$GHz case shown in the bottom panel, scattering is no longer an important timing error, allowing $x_\perp \approx 25\,$AU  to achieve $\sigma_d/d \approx 0.1$ out past $d=1000\,$Mpc for $\delta x = 1~$cm, with a 1\% measurement to $d=200\,$Mpc.  We emphasize that these forecasts are for a \emph{single} FRB.  

Again considering our fiducial $\delta x$ and moving to a constellation with $x_\perp \approx 50\,$AU (the green curves), while the explanation for the trends are the same, at $4\,$GHz now 10\% measurements are possible out to $d= 1000\,$Mpc, with better than a $2\%$ precision for $d = 200~$Mpc.  At $8\,$GHz with again $x_\perp \approx 50\,$AU, 10\% measurements are possible out to either 1000\,Mpc and $10000$~Mpc, depending on whether Shapiro delays are not or are removed.  At $x_\perp \approx 100\,$AU (the red curves), the errors at small $d$ improve by yet another factor of $(50/100)^2$ owing to the larger $\Delta t_d$, but cosmological Shapiro delay is starting to become the limiting factor at large $d$ for the solid curves.  Finally, for $x_\perp \approx 200\,$AU, Shapiro delay cleaning is becoming essential as there is little improvement over the $x_\perp \approx 100\,$AU case without cleaning.  Once the Shapiro delay is removed (as for the dashed curves), then sub-percent measurements to $10^4~$Mpc ($z\approx 5$) in the concordance cosmology are possible at $8\,$GHz.  Thus, the scaling $\Delta t_d \propto x_\perp^2/d$ results in a fast increase in the distance reach with increased average baseline length.

The dotted curve in Figure \ref{fig:distance_uncertainty} is the fractional redshift uncertainty that arises if the peculiar velocity contribution to the redshift can be corrected using large-scale structure observations to $\delta v = 200\,$km\,s$^{-1}$, \tmp{about half of the $z=0$ RMS value for the line-of-sight component and in line with the claimed removal in cosmological analyses of Type 1a supernovae \citep{Peterson2022}.}  The fractional redshift uncertainty is calculated as $\delta v \times (1+z)/[c\,z]$. While for certain configurations the distance can be constrained to higher precision, there is little gain in doing so below this peculiar velocity limit. 

All of the previous results are on a per-source and per-burst basis.  Individual Type Ia supernova distances only constrain distance to $\approx 10\%$, requiring about $100$ supernova for the current 1\% measurement of $H_0$ \citep{2021arXiv211204510R}.  Analogous statistical improvements using multiple FRBs can of course be done here.  \tmp{Furthermore, repeated observations of repeating FRBs will offer a consistency check, especially since the scattering time and differential dispersion to each detector change on month to year timescales.}


\section{Other potential science}
\label{sec:otherscience}

Additionally, we have identified several other potential sciences to which the proposed experiment could contribute:
\begin{description}
    \item[the mass distribution in the outer Solar System]  The unaccounted gravitational accelerations of the satellites that lead to $\gtrsim 1\,$cm displacements owes to asteroids with radii of $R\gtrsim 10 \,(v/10 \,\text{km s}^{-1})^{2/3}~$km (\S~\ref{sec:calibration}).  Such displacements are potentially detectable if non-gravitational accelerations can be isolated.  Additionally, in a toy picture where mass is spherically distributed, an excess mass of just $10^{-7}(r/100 \,{\rm AU})^{-1/2}M_\earth$ interior to an orbiting satellite results in a potentially measurable centimeter displacement in one year.   Similarly, this experiment may also be sensitive to dark matter clumps with masses $\gtrsim (200\, \text{km/s})^2 
    \times 1\text{cm}/G =  10^{-6}\,M_\earth$ (similar to the anticipated dark matter mass within a $100\,$AU cube in the Solar neighborhood), which some dark matter models predict should be passing near the Solar System with typically a much larger velocity of $\sim 200\,$km~s$^{-1}$ compared to bound outer Solar System bodies.
     \item[geometric distances to and radio emission from Milky Way pulsars]  As our system is capable of measuring distances out to hundreds of megaparsecs, from a geometric standpoint it can easily measure distances (even with $0.1~$AU baselines) to sufficiently bright pulsars at all distances in the Milky Way, \tmp{with the caveat that individual pulses from most pulsars are difficult to detect with 10~m dishes especially at the higher frequencies at which the pulsars have $\lesssim \Delta t_d$ scattering times as would be required.  
     Such precise distances would enhance the gravitational wave sensitivity of pulsar timing arrays \citep{2012PhRvD..86l4028B} and even enable pulsar timing arrays to measure $H_0$ \citep{2022MNRAS.517.1242M}.} 
     Furthermore, the proposed experiment could shed light on the nature of pulsar emission.  It could resolve their radio emission, as a baseline's spatial resolution is $\ell =  10^2\, \text{km}~(x_\perp/100 \,\text{AU})^{-1} ( \nu/5 \,\text{GHz})^{-1} (d/1 ~\text{kpc})$.  The nature of pulsar radio emission is still debated, with two classes of models \citep[e.g.][]{doi:10.1146/annurev-astro-052920-112338}.  The first predicts that the emission is from the inner magnetosphere at radii of $10-100~$km.  A $100~$AU baseline has the potential to resolve this component, especially for the closest pulsars. The other class of emission occurs at the $\sim 5000 \,(\tau/0.1\,s)\,$km light cylinder, where $\tau$ is the spin period, which would be resolved with $x_\perp \gtrsim 10 \,(\tau/0.1~s)^{-1} (\nu/1\,\text{GHz})^{-1}~$AU. 
\item[the spectrum of density fluctuations from both ISM electrons and the dark matter]  The dispersion delay differences between different satellites should be easily detectable (\S~\ref{sec:dispersion}). Such measurements would constrain the spectrum of ISM density fluctuations on the Solar System-scale of our antenna separations, complementing measurements on smaller and larger scales.  \tmp{This would be interesting for measuring the spectrum of turbulence \citep{1981Natur.291..561A} and also for testing the picture of electron inhomogeneities from $\sim 1~$AU current sheets invoked to explain some aspects of scattering and scintillation \citep{2006ApJ...640L.159G, 2014MNRAS.442.3338P}.}  For some sightlines the ISM scattering to each antenna would also be constrained, further testing these models.  Our calculations also suggest that a six satellite system with $x\gtrsim 100~$AU would measure the quadrupolar moments of the Shapiro time delay.  In the standard cosmology, these moments are most sensitive to the abundance and profiles of $10^{10}-10^{13}M_\odot$ dark matter halos, but they can be shaped by much smaller structures in clumpier dark matter models (Appendix~\ref{ap:shapiro}).   
\item[micro-Hertz gravitational waves] Timing signals between satellites is sensitive to gravitational wave strains with frequencies $\lesssim 10^{-4}(x/20\,\text{AU})^{-1}$~Hz. Strains of $10^{-18}- 10^{-16}$ around such frequencies are predicted from various Galactic and cosmological sources \citep{2021ExA....51.1333S}.  While a strong science case exists for the `midband' frequencies between the nano-Hertz frequencies measured by pulsar timing and the $10^{-4}-1\,$Hertz frequencies probed by LISA \citep{2017arXiv170200786A}, they unfortunately are difficult to target with the traditional laser ranging in the inner Solar System \citep{2022PhRvD.105j3018F}. At our long baselines, radio ranging is likely the more practical approach because the SNR in the match-filtered electric field scales inversely with the radiometer separation compared to the quadratic fall off in SNR because of the dilution of laser photons (and the laser power requirements are already challenging for 1\,AU baselines and reasonable mirror sizes; \citealt{2021ExA....51.1333S, 2021PhRvD.103j3017F}).  Here we consider detecting gravitational waves by measuring baseline distances rather than the less-ambitious Doppler ranging that has been applied to outer Solar System spacecraft \citep{dopplerranging}. How well the carrier wave phase can be timed between two of our satellites is limited by the frequency stability of the onboard atomic clock on at least one satellite (as the signal from other satellites can be phased to the incoming signal), with a stability characteristic of the Deep Space Atomic Clock allowing sensitivity to millimeter displacements for gravitational waves with strain errors of $\delta h=10^{-15} (\nu_{\rm GW}/10^{-5}~\text{Hz})^{-1/2} (x/100\, {\rm AU})^{-1} W^{-1}$, where $W$ is a window function that is near unity for gravitational wave frequencies of $\nu_{\rm GW} \sim c/x = 2\times 10^{-5} (x/100\, {\rm AU})^{-1}\;$Hz and scales as $W\sim \nu_{\rm GW} x/c$ at lower $\nu_{\rm GW}$ (since $x$ spans only a fraction of a wavelength).  The best atomic clocks in the world have been improving rapidly and would best the timing precision of the Deep Space Atomic Clock by five orders of magnitude. Once no longer limited by the onboard clock, the precision that the carrier wave can be timed as well as stochastic sources of gravitational acceleration likely set the error.  The precision that the gravitational wave strain can be measured when limited by timing of the carrier phase to accuracy $\delta t_c$ is \citep{misraenge} 
\begin{eqnarray}
\delta h =\frac{c \, \delta \, t_c}{x}\, W^{-1} &\sim& \frac{c}{2\pi \nu x}\sqrt{\frac{\nu_{\rm GW} k T_{\rm sys}}{ P_{\rm rec}}} \, W^{-1},\\ 
&=&   2\times 10^{-20} \, W^{-1} \left(\frac{5\,\textrm{GHz}}{\nu}\right)^2    \left(\frac{10\,\textrm{m}}{D}\right)^{2}  
    \sqrt{ \frac{\nu_{\rm GW}}{10^{-5} \,\text{Hz}} \frac{100\,\text{Watt}}{P_\text{ant}}  \frac{T_\text{sys}}{20\,\text{K}}},
\end{eqnarray}    
    where this assumes continuous tracking over the full wave period.  Gravitational accelerations on the orbital timescales of bodies likely sets the redward limit to $\nu_{\rm GW} \gtrsim 0.01\,\mu$Hz \tmp{assuming that the non-gravitational accelerations can be isolated}.   
\end{description}

\section{Conclusions}
\label{sec:conclusions}

We have described a new method for measuring cosmological distances using the differences in
arrival times to cosmological sources as measured by extremely long baseline interferometry.  In order to vastly improve future measurements of the late-time expansion of the Universe, we think that the proposed experiment may be our best option.  Despite the huge number of potentially complicating effects, the proposed experiment seems possible for a surprisingly well-defined set of mission specifications:
\begin{itemize}
    \item FRBs are the one source class that is bright enough and sufficiently compact for this measurement. 
    \item  \tmp{We showed that once baselines reach $x =25$\,AU, individual FRBs can be used to measure percent-level distances to $d=200\,$Mpc ($z=0.05$), and a ten percent measurement to $d=2000\,$Mpc  ($z=0.5$).
    Going up in scale to $x\approx 100~$AU, 1\% measurements on the distance are possible for a $z=1$ FRB.  In the limit where the detector space-time positions are known, the primary limitation on the timing precision for baselines of $x <100\,$AU is Milky Way ISM scattering at $\lesssim 3\,$GHz and differential dispersion-marginalized VLBI timing errors at higher frequencies, resulting in a timing noise of $\sigma_t \sim \nu^{-1}$ at $\nu \gtrsim 3\,$GHz.}     
    \item  The experiment is best at $\nu\gtrsim 3\,$GHz to minimize Milky Way scattering. A wide bandwidth with $\Delta \nu/\nu \gtrsim 0.1$ (or multiple bands) is helpful for signal-to-noise considerations and to subtract relative delays from dispersion based on their frequency dependence, delays we find can be comparable to the geometric delay of interest.   The requirement to use a higher frequency band than the $\nu \sim 1\,$GHz targeted by most FRB surveys is perhaps the biggest uncertainty in this proposal, as the FRB population is less studied there. For baselines of $x\gtrsim 100\,$AU, the curvature in the total Shapiro time delay limits the precision, and we showed the cosmological contribution to this delay is typically larger than the Milky Way contribution from stars.  The quadrupolar shape of the contaminating Shapiro delay allows it to be removed at the cost of two additional satellites.
    \item The satellite positions need to be determined to several centimeters for our forecasts to apply, with the error increasing linearly with the positional uncertainty once above this specification.  We showed that GNSS-like direct trilateration using the dishes themselves appears feasible, although calibration may also be possible using only distant FRBs themselves.  The latter calibration strategy likely requires a precise on-board accelerometer to eliminate non-gravitational accelerations and, thus, extend the timescale needed between calibrations.  We estimate that gravitational accelerations from closely passing asteroids should lead to a $1\,$cm error in satellite ephemeris over a time of months at the very minimum, but likely much longer.  Without a precise accelerometer, ephemeris calibration must be performed weekly to account for accelerations from variations in the Solar irradiance, dust collisions, gaseous drag, and possibly Lorentz forces acting on the satellites.  For weekly calibration, the satellites' clock accuracy requirement is in line with the demonstrated accuracy of NASA's Deep Space Atomic Clock.
    \item The increasing catalog of FRB sources known to repeat provides a list of targets for which many are likely to yield an FRB after just hours of integration.  The simplest approach would be for the satellites to point to such a source and store the voltage time series, telemetering it back to Earth if prompted that a burst occurred. A more ambitious approach that would require substantial onboard computation is for each satellite to search the voltages and identify FRBs on the fly. Then, 10\,m diameter dishes are necessary with tens of beams to detect tens of FRBs a year, although there is substantial uncertainty in how the observed $\sim 1\,$GHz FRB rates extrapolate to $\gtrsim 3\,$GHz.  Possibilities that could greatly increase the detection rates are (1) using a more sensitive terrestrial radio telescope for a template of the electromagnetic waveform, and (2) employing phased elements in each satellite rather than a single dish to simultaneously observe much of the sky.  A 10\,m diameter instrument is also convenient for ephemeris calibration by direct trilateration, requiring only $\sim 10\,$Watt broadcasts between antennas separated by $100\,$AU for $\sim 0.1~$ns timing using a modulating $10\,$MHz code and minute integration times.  A fold-out design (as done for the $D=10\,$m Spektr-R satellite that had a 30\,$R_\earth$ apogee) could allow a radio dish to be deployed in a modest payload. 
    \item We favor unbound orbits where the baselines gradually drift to larger and larger separations.  \tmp{As an example, the New Horizons spacecraft achieved a terminal velocity of $10\,$km~s$^{-1}$ via gravity assists off of planets, enabling it to travel $40~$AU in a decade.  Unbound orbits would allow nailing down science that can be accomplished at shorter baselines first and a quadratic-with-time increase in the size of $\Delta t_d$, and hence the precision of distance determinations, as the detectors drift apart.  It might even be possible to launch all the antennas in one rocket and disperse them in different directions by scattering off a Solar System body.}
\end{itemize}

Some studies have discussed the possibility of detecting the parallaxes of cosmological objects (e.g.~\citealt{2009MNRAS.397.1739D, 2022arXiv220305924C}).  The proposed experiment to detect wave front curvature can be thought of in terms of parallax with instruments separated by $\sim x_\perp$ and where the angular resolution is set by $\sim \lambda/x_\perp$ (\S~\ref{sec:geometric_dist}).  If we take our fiducial specifications of $x_\perp =100~$AU and $\nu = 5$~GHz, such an experiment's angular resolution is more than five orders of magnitude more precise than the state-of-the-art $\sim 100~\mu$-arcsecond astrometric localizations that the Gaia satellite achieves toward the brightest quasars \citep{2016A&A...595A...2G}.  Hence the effective sensitivity of our method for measuring distance is improved by a similar factor (times the additional factor of $\sim x_\perp/(2~\text{AU})$ when comparing to annual parallax).\\




\tmp{We also identified several other interesting sciences that could be accomplished with the proposed instrument (\S \ref{sec:otherscience}).
These include measuring the distance and resolving the radio emission region of Galactic pulsars, constraining the mass distribution in the outer Solar System at the millionths of Earth mass-level (interesting both for Solar System and dark matter science), direct constraints on the density distribution in the interstellar medium (ISM) on $x_\perp$ scales, and potentially interesting sensitivities to $\sim 0.01- 100\, \mu $Hz gravitational waves.} The proposed instrument not only would provide a more direct means to constrain distance in cosmology, but also contribute to answering other key questions.
    
\begin{acknowledgments}

We thank Yakov Faerman, Jason Hessels, Bryna Hazelton, Leonid Gurvits, Kiyoshi Masui, Casey McGrath, Miguel Morales, Ue-Li Pen,
Alexander Philippov, Alexander Tchekhovskoy, and Huangyu Xiao for useful discussions and comments.  We thank the two anonymous referees for valuable feedback. MM acknowledges support from NSF award AST-2007012.  KB acknowledges support from the DiRAC Institute in the Department
of Astronomy at the University of Washington. The DiRAC Institute is supported
through generous gifts from the Charles and Lisa Simonyi Fund for Arts and Sciences and from the Washington Research Foundation.

\software{
    Astropy \citep{astropy13, astropy18},
    emcee \citep{foremanmackey13},
    Matplotlib \citep{hunter07},
    NumPy \citep{numpy},
}

\end{acknowledgments}

\appendix
\section{FRW spacetime derivation}
\label{sec:curvedspace}
In flat space, the distances that appear in the main text are really the effective distance from the accumulated phase in the wave fronts. Namely,
\begin{equation}
    d = \lambda_{\rm obs} \int_{t_{\rm em}}^{t_{ \rm obs}} \frac{c \, dt}{\lambda} = \int_0^z \frac{c \, dz}{H(z)},
    \label{eqn:lighttraveldistance}
\end{equation}
and similarly for $d_x$ (c.f.~Figure~\ref{fig:detector_geometry}), 
where $t_{\rm em}$ and $t_{ \rm obs}$ are, respectively, the emission and observation times and $\lambda = \lambda_{\rm obs}/(1+z)$ relates the wavelength to its observed (redshifted) value.  These reduce to the Euclidean distance at low redshift. The rightmost integral in Equation~\ref{eqn:lighttraveldistance} is just the comoving light-travel distance.  Thus, one is free to draw the antenna--antenna--FRB triangle in Figure~\ref{fig:detector_geometry} at the present time, ignoring the time dependence of FRW expansion.  The same result is well established for lensing time delays \citep{1996astro.ph..6001N}.

To generalize our calculation to curved FRW space-times, Figure~\ref{fig:detector_geometry} should be redrawn as a triangle in an open or closed geometry and, rather than using the law of cosines to calculate $d_x$ (c.f. Equation~\ref{eqn:lawofcosines}), we use these spaces' generalization: 
\begin{align}
    \cos(\theta_{d_x}) & = \cos(\theta_d)\cos(\theta_{x})+ \sin(\theta_d)\sin(\theta_{x}) \cos(\angle_{d-x}),\\
    & \approx  \cos(\theta_d)(1 - \theta_{x}^2/2)+ \sin(\theta_d)\theta_{x} \cos(\angle_{d-x}),
    \label{eqn:sphericallawofcosines}
\end{align}
where $\angle_{d-x}$ denotes the angle adjoining the line segments $d$ and $x$ in the triangle and $\theta_r \equiv k r/R$, with $r$ as a stand-in for other distance variables.  The value $k=1$ corresponds to a closed geometry and $k=i$ to an open one, where $R^{-1} \equiv \sqrt{|\Omega_k|} H_0/c$ is the present-day radius of curvature.\footnote{These relations for the law of cosines can be found on Wikipedia, with references dating into the nineteenth century.  The identification with the curvature space density $\Omega_k$ for FRW cosmologies is given in many cosmology textbooks \citep[e.g.][]{peacock}.}  Equation~\ref{eqn:sphericallawofcosines} takes the applicable limit that $\theta_{x}$ is much less than one.  To further simplify, we note that $ \cos(\theta_{d_x})  \approx \cos(\theta_d) - \sin(\theta_d) k (d_x - d)/R -  \cos(\theta_d) [k (d_x - d)/R]^2/2$.  
An equation quadratic in $d_x -d$ results, which solving yields the root: 
\begin{eqnarray}
c \,\Delta t &=& d_x -d  =    \frac{R}{ k}\left[- \tan(\theta_d) + \sqrt{\tan(\theta_d)^2 +  \frac{ k^2 x^2}{R^2} - 2 \tan(\theta_d)\frac{k x}{R}  \cos(\angle_{d-x} )} \right],\\
&\approx& -  x \cos(\angle_{d-x} ) + \frac{k x^2 \sin(\angle_{d-x} )^2}{2 R \tan(\theta_d)}  + {\cal O}\left (\frac{x^3}{R^2\tan(\theta_d)^2} \right),\\
&=& - \bfx \cdot \hatd + \frac{k x_\perp^2}{2 R \tan(\theta_d)}  + {\cal O}\left (\frac{x^3}{R^2\tan(\theta_d)^2} \right),\label{eqn:full}
\end{eqnarray}
where the second line reapplies the limit $x/R \ll 1$ and $x_\perp \equiv x \sin(\angle_{d-x} )$.  Equation~\ref{eqn:full} shows that a measurement of the time delay between local detectors is sensitive to the cosmological distance $R \tan(\theta_d)$.  This distance is similar to the luminosity and angular diameter distances that many cosmological observables are sensitive to, which are given by $R \sin(\theta_d)$ up to factors of $(1+z)$.

In conclusion, the curved-space generalization of the time delay owing to curvature of the wave front is
\begin{equation}
\Delta t_d  \approx   \frac{k x_\perp^2}{2 c R \tan(\theta_d)} .
\end{equation}
In the limit that the space's radius of curvature $R$ is much larger than $d$, this reduces to our Euclidean result in \S~\ref{sec:geometric_dist} but with Equation~\ref{eqn:lighttraveldistance} for the distance.

We learned after submitting this manuscript to the preprint server that a nearly identical derivation as above was done in \citet{2022MNRAS.517.1242M}, considering the effect of curvature of the wave front of gravitational waves on pulsar timing arrays.  Furthermore, \citet{1972gcpa.book.....W} define the distance $\tan(\theta_d)$ as the parallax distance.  Since wave front timing can be thought of in terms of parallax (\S~\ref{sec:geometric_dist}), that wave front curvature probes the parallax distance is not surprising.

\section{fitting for the dispersion}
\label{ap:fittingdisp}

 \tmp{ Here we calculate how well the effects of dispersion to each detector can be fit for and removed.   We again use the notation that $\Delta t$ is our parameter for the geometric delay, and we adopt $\Delta \tau_d$ their differential dispersive delay.  We want to find the delays $\boldsymbol{p}= (\Delta t, ~\Delta \tau_d)$ that  minimize $\chi^2 = \sum_{i} | {\cal V}_{i} - {\cal V}_{i, \rm obs}|^2/\sigma_i^{2}$, where ${\cal V}_i \equiv  \langle E_{1i} E_{2i}^* \rangle$ is the complex correlation of the electric field in frequency channel $\nu_i$ between two detectors (the `visibility') computed by sampling different times across the FRB, the subscript `obs' indicates the observed, whereas without it is the modeled, and $\sigma_i$ is the channel noise on ${\cal V}_{i, \rm obs}$.   Critically, since a time delay results in a phase in Fourier space, delays alter ${\cal V}_i$ by giving it a complex phase of $\exp[i   2\pi \nu_i (\Delta t + \Delta \tau_d (\nu_i/\bar \nu)^{-2})]$, where $\bar \nu$ is the SNR-weighted midpoint of the band. The Fisher information matrix (which describes the curvature of the log likelihood) is given by $F_{nm} = 1/2 \, d\chi^2/dp_n dp_m \approx \sum_i \sigma_i^{-2} d {\cal V}_i/dp_n d{\cal V}_i^*/dp_m$ \citep[e.g.][]{2003moco.book.....D}, and the $1\,\sigma$ uncertainty on each delay parameter is given for $\Delta t$ by $\boldsymbol{F}^{-1/2}|_{\Delta t \Delta t}$.  (We do not need to include the amplitudes of the visibility as another parameter because they are uncorrelated with the other parameters -- $\boldsymbol{F}$ is diagonal in this parameter -- and so can be fit independently.)}

\tmp{  First let us consider the case where we do not know the absolute phase $\theta_i$, defined as the phase at the frequency midpoint of the band $\bar \nu$.   The absolute phase can be complicated by clock phase errors -- and for terrestrial measurements by the atmosphere -- and also the space-time position of the detectors must be known to better than a wavelength for solutions for $\theta_i$ to not have a phase-wrapping degeneracy.  In this case, we rewrite the complex visibility phase as $\exp[i \{\theta_i + 2\pi (\nu_i -\bar \nu)  (\Delta t + \Delta \tau_d (\nu_i/\bar \nu)^{-2}) \} ]$.   As $\Delta t$ only includes the phase across the band and the rest is absorbed into $\theta_i$, this case is measuring the equivalent of the group delay over the band ($d \phi/d\nu$ when $ \Delta \tau_d = 0$).  This yields the VLBI group delay timing noise $\sigma_{t, \rm group}^{\rm VLBI}$ given by Equation~\ref{eqn:sigmat} in the case where differential dispersion is not a parameter so that $F$ is a scalar. (We do not need to consider $\theta_i$ as a parameter as referencing $\Delta t$ to the effective center of the band makes this parameter independent; see Appendix 12.1 in \citet{ 2017isra.book.....T}.)  One can check that it does, noting that $d {\cal V}_i/d\Delta t  \approx 2\pi i (\nu - \bar \nu)\, {\cal V}_i$.   If we then sum over all frequency channels assuming the signal to noise is channel independent, this yields $F = [2\pi \, \Delta \nu_{\rm RMS} {\rm SNR}]^2$, where ${\rm SNR}^2 = \sum_i |{\cal V}_i|^2/\sigma_i^2$ is the total signal-to-noise ratio on the intensity that the baseline measures.  This agrees with Equation~\ref{eqn:sigmat}, except there we wrote there the more general expression for when the channel signal-to-noise ratio varies with frequency.  In other references this result is derived by least squares fitting directly to the phase \citep[e.g.][]{1970RaSc....5.1239R, 2017isra.book.....T}.}

\tmp{Now including dispersion in this group velocity limit where we are only using the phase trends across the band, this becomes a three parameter Fisher matrix calculation as $\theta_i$ is weakly correlated with $\Delta \tau_d$.  In this case, the error becomes significantly larger,  $\approx 10 \, \sigma_{t,\,\rm group}^{\rm VLBI} (0.2\,\nu/\Delta \nu )$.  For $\Delta \nu/\nu \gtrsim 0.1$, interesting cosmological constraints may still be possible with such timing noises (\S~\ref{sec:discussion}).}

\tmp{However, there is no reason to throw away the absolute phase information as the design of the proposed experiment requires precise baseline measurements.  When using the full phase, the Fisher matrix becomes  $\boldsymbol{F} \propto [\int_{\Delta \nu} \nu^2 \, d\nu ~~~  \bar \nu^2 \Delta \nu;  ~\bar \nu^2 \Delta \nu ~~ \bar \nu^4 \int_{\Delta \nu} \nu^{-2}\, d\nu]$.  One can show that the error on the dispersion-marginalized phase velocity is exactly $\sqrt{[\boldsymbol{F}^{-1}]_{0,0}} =  \sigma_{t, \rm group}^{\rm VLBI}/2 = [4 \pi \Delta \nu_{\rm RMS} {\rm SNR}]^{-1} $. The delay from the absolute phase is subject to a $2\pi$ phase-wrapping degeneracy.  The correct phase can be distinguished with multiple frequencies or a broad band.  The $n^{\rm th}$ phase fringe can be distinguished from the zeroth at $1\sigma$ once $\nu/\Delta \nu \lesssim  n \,  {\rm SNR}$.}

Global navigation systems often use a second frequency band to remove ionospheric delays.  Their signal template is perfectly known, in contrast to the noisy `template' supplied by the second VLBI receiver.  However, the results carry over: Our dispersion-marginalized noise estimate in the narrow two-channel limit increases the timing noise by the factor $(\nu_1^2+ \nu_2^2)/(\nu_1^2- \nu_2^2)$ over the timing noise in either channel in the absence of dispersion, reproducing the result known for two-channel GNSS \citep{misraenge}.

\tmp{We find that simultaneously fitting for a term $\propto \nu^{-4}$ to remove weak refractive scattering (\S~\ref{sec:scattering}) comes at a larger cost. 
When using the absolute phase, the error is increased to $\approx 5 \,\sigma_{t, \rm group}^{\rm VLBI} (\Delta \nu /0.2 \,\nu)$.  However, since we favor targeting frequencies where the scattering time delays are $\lesssim \nu^{-1}$, such marginalization is unnecessary to reach the timing goal of $\sigma_t \sim \nu^{-1}$.}

\tmp{The above assumes that the dispersion is fit for each visibility independently.  One can also fit for the geometric delay $\Delta t_d$ and the $N_A - 1$ differential dispersion delays simultaneously.  This approach incurs a similar cost for marginalizing out dispersion, since dispersion is distinguished from geometric time delays via its frequency dependence.}

\section{Dispersion delays from ISM turbulence}
\label{sec:dispturb}
Here we estimate the delays from ISM turbulence leading to varying electron columns along the sightlines to the different detectors.  The variance of the differences in dispersive delays $\tau_d$ between pairs of sightlines is given by
\begin{align}
    \textrm{Var}\left[\tau_{d,1} - \tau_{d,2}\right] &= 2 \expectation{\Delta\tau_{d,1}^2} - 2 \expectation{\Delta\tau_{d,1} \Delta\tau_{d,2}},
    \label{eq:dt_dispersion_var}
\end{align}
where we write the difference in the dispersion time delay of the sightline with path $P_i$ relative to the mean value as
$\Delta \tau_{d, i}$, angular brackets indicate an ensemble average, and
\begin{align}
    \expectation{\Delta\tau_{d,1} \Delta\tau_{d,2}} 
     &= \kappa^2 \expectation{\int_{P_1} dx_1 \Delta n_e(\bfx_1) \int_{P_2} dx_2 \Delta n_e(\bfx_2)},
\end{align}
 and $\Delta n_e$ is the 3D field of electron density fluctuations. Replacing $\Delta n_e$ with its Fourier transform, the expectation value becomes
\begin{align}
    \expectation{\Delta\tau_{d,1} \Delta\tau_{d,2}} = & \kappa^2 \int_{P_1} dx_1 \int_{P_2} dx_2 \iint \frac{d^3k_1 d^3k_2}{(2\pi)^6}  \expectation{\Delta n_e(\bfk_1) \Delta n_e(\bfk_2)^*} e^{-i (\bfk_1 \cdot \bfx_1 - \bfk_2 \cdot \bfx_2)}, \\
    = & \kappa^2 \int_{P_1} dx_1 \int_{P_2} dx_2 \int \frac{d^3k}{(2\pi)^3} \nonumber  P_e(k) e^{-i \bfk \cdot (\bfx_1 - \bfx_2)},
\end{align}
where we have used the property of real fields that $\Delta n_e(\bfk)^*  = \Delta n_e(-\bfk)$ and defined the electron density power spectrum as
    $\expectation{\Delta n_e(\bfk_1) \Delta n_e(\bfk_2)^*} = (2 \pi)^3 P_e(k_1) \delta^D(\bfk_1-\bfk_2)$.
Evaluating the integral over the line-of-sight wavenumber along $P_1$ gives a $\delta$-function that eliminates the light of sight wavenumber integral.
We next evaluate the spatial line of sight integral along $P_2$ to a length of $L$, representing the size of the region containing
electron density fluctuations (i.e. the extent of the Milky Way ISM), yielding
\begin{align}
    \expectation{\Delta\tau_{d,1} \Delta\tau_{d,2}} = \kappa^2 L \int \frac{d^2k_\perp}{(2\pi)^2} P_e(k) e^{-i k_\perp \cdot (\bfx_{\perp,1} - \bfx_{\perp,2})} = \kappa^2 L\int_0^\infty \frac{dk}{2 \pi} k P_e(k) J_0(k x_\perp),
        \label{eqn:vartauintegral}
\end{align}
where the last equality assumed that the paths are parallel and separated by a transverse distance of $x_\perp$. 
%
We can use this expression to evaluate the square of the standard deviation in dispersion time delays between the two lines of sight defined in Equation~\ref{eq:dt_dispersion_var}:
\begin{align}
    \sigma^2\left[\tau_{d,1} - \tau_{d,2}\right] &= 2 \kappa^2 L \int_0^\infty \frac{dk}{2 \pi} k \, P_e(k)\, \left[1 - J_0(k x_\perp)\right].
\end{align}
For the Milky Way, the electron density distribution is given by a Kolmogorov-like power law $P_e(k) = (2\pi)^3 C_n^2\, k^{-11/3}$ between
some $\sim 10^9\,$cm inner and $\sim 1\,$pc outer length scales, where the $(2\pi)^3$ owes to our Fourier convention as we have adopted the standard definition for $C_n^2$ \citep[e.g.][]{draine11}. In the solar neighborhood, measurements find $C_n^2 \sim 5 \times 10^{-17} \textrm{cm}^{-20/3}$ \citep{1981Natur.291..561A, draine11}. This integral is not sensitive to the inner or outer length scales, and evaluating the square root yields the standard deviation
\begin{align}
    \sigma\left[\tau_{d,1} - \tau_{d,2}\right] &= 9.4~\kappa \, C_n x_\perp^{5/6} \sqrt{L},\\ 
    &= 280~\textrm{ns}
    \left(\frac{L}{0.1~\textrm{kpc}}\right)^{1/2}
    \left(\frac{x_\perp}{100~\textrm{AU}}\right)^{5/6}
    \left(\frac{\nu}{5~\textrm{GHz}}\right)^{-2} 
\left( \frac{C_n^2}{5 \times 10^{-17} \textrm{cm}^{-20/3}}\right)^{1/2}.
    \label{eqn:vartau}
\end{align}
Much of this contribution to $\sigma\left[\tau_{d,1} - \tau_{d,2}\right]$ is from long-wavelength modes that manifest as a gradient that is degenerate with the $-\bfx\cdot \hatd$ delay from the source location on the sky.  Crudely, only the contribution of modes in $k > \pi/x_\perp$ could bias the delay from wave front curvature.  These wavenumbers (and indeed the same holds when restricting to the even more curvature-like modes with $\pi/x_\perp < k < 2\pi/x_\perp$) contribute only 10\% of the integral in Equation~\ref{eqn:vartauintegral}.  Thus, the fraction that is non-planar and, hence, contaminates our distance measurement is
\begin{align}
    \sigma\left[\tau_{d,1} - \tau_{d,2}\right]_{\rm non-planar} \approx 90~\textrm{ns}
    \left(\frac{L}{0.1~\textrm{kpc}}\right)^{1/2}
    \left(\frac{x_\perp}{100~\textrm{AU}}\right)^{5/6}
    \left(\frac{\nu}{5~\textrm{GHz}}\right)^{-2} \left( \frac{C_n^2}{5 \times 10^{-17} \textrm{cm}^{-20/3}}\right)^{1/2}.
    \label{eqn:sigmadispap}
\end{align}
We use this expression for $\sigma_t^{\rm disp}$ in \S~\ref{sec:dispersion} (e.g. eqn.~\ref{eqn:sigmadisp}).

\section{Cosmological contribution to the Shapiro delay}
\label{ap:shapiro}

This appendix calculates in detail the cosmological Shapiro time delay from large-scale structure.  As the Shapiro delay along a sightline can be written as $\frac{2}{c^3}\int_0^{d} d\chi \Phi$, where $\Phi$ is the Newtonian gravitational potential, the Shapiro delay between two sightlines separated by $\bfx_\perp$ can be calculated via Taylor expansion assuming the potential is smooth on the scale $\bfx_\perp$: $\Delta t_{\rm grav} = \frac{2}{c^3} \int_0^d d\chi ({\bfx_\perp}_i  \bfnabla_j \Phi + \frac{1}{2} {\bfx_\perp}_i {\bfx_\perp}_j \nabla_i \nabla_j \Phi +...)$.  The quadratic-in-$x_\perp$ term inside the integral can be rewritten as $\frac{1}{2} {\bfx_\perp}_i {\bfx_\perp}_j \frac{\nabla_i \nabla_j}{\nabla^2} 4\pi G \rho$ using Poisson's equation for the Newtonian gravitational potential and noting that only long-wavelength modes in the line-of-sight direction contribute and so we can ignore the line-of-sight derivatives. We can write the variance of the cosmological component of $\Delta t_\text{grav}^\text{quad}$ in terms of the matter overdensity power spectrum using an approach nearly identical to in \S~\ref{sec:dispturb} except the different angular weighting that appears:
\begin{eqnarray}
      \text{Var}[\Delta t_\text{grav}^\text{quad}] &=&  \frac{(4\pi G \bar \rho)^2 d}{c^6} \int \frac{d^2\bfk_\perp }{(2\pi)^2} \,  \frac{(\bfk_\perp \cdot \bfx_\perp)^4}{k_\perp^4} P_\delta(k_\perp) \exp[-i \bfk_\perp \cdot \bfx_\perp], \\
            & = &
         \frac{(4\pi G \bar \rho)^2 d\, x_\perp^4}{c^6} \int \frac{d k_\perp }{(2\pi)} J_2(k_\perp x_\perp)  k_\perp \left( \frac{3}{(k_\perp x_\perp)^2} -1 \right)   \,  P_\delta(k_\perp),\\
         &\approx&   \frac{3(4\pi G \bar \rho)^2 d \,x_\perp^4}{8c^6}  \int \frac{d k_\perp }{(2\pi)} k_\perp P_\delta(k_\perp),\label{eqn:cosmoshapfinal}
\end{eqnarray}
where we have further assumed no evolution in the statistics of the density field over the sightline (which should make this inaccurate for $d\gtrsim 3000\;$Mpc, corresponding to $z\gtrsim 1$) and the last approximation uses that the contribution to the integral owes to structures with $x_\perp k_\perp \ll 1$, at least in the standard cosmology.  To evaluate the integral in Equation~\ref{eqn:cosmoshapfinal}, we adopt the $z=0.1$ HaloFit model for the nonlinear matter power spectrum \citep{2003MNRAS.341.1311S}, finding that the standard deviation of the cosmological Shapiro time delay is
\begin{equation}
    \sigma_\text{t, \text{grav}}^{\rm cosmo} \approx 0.08 ~{\rm ns}~ \left(\frac{x_\perp}{100~\textrm{AU}}\right)^2 \left(\frac{d}{100~\text{Mpc}}\right)^{1/2}.
    \label{eqn:tgravquadcosmo2}
\end{equation}
This is somewhat larger than our estimate for the delay from stars in the Galactic disk.  A significant fraction of this variance can be reduced by selecting sightlines that do not pass near large dark matter halos. (Half of the contribution to this integral comes from halo virial radius scales of $k \in 2\pi/[0.1-1~\text{Mpc}]$, and for sightlines with a significantly enhanced cosmological Shapiro delay the contribution from large halos would be even larger.) However, measuring this delay would itself be interesting as it is a probe of nonlinear scales, with an identical wavenumber weighting of modes as the RMS magnification from gravitational lensing, and so constraining $\sigma_\text{t, \text{grav}}^{\rm cosmo}$ would be very analogous to the supernova magnification constraints that have placed limits on the clumpiness of the dark matter \citep{2006PhRvD..74f3515D, 2018PhRvL.121n1101Z}.  

\bibliography{references}{}
\bibliographystyle{aasjournal}



\end{document}